\documentclass[a4paper,11pt]{article}
\pdfoutput=1 
\usepackage{jcappub} 
\usepackage{commath}
\usepackage{mathtools}
\usepackage{float}
\usepackage{url}
\usepackage{tensor}
\usepackage{esdiff}
\usepackage{tabularx}
\usepackage[dvipsnames]{xcolor}
\usepackage{caption}
\usepackage{subcaption}
\usepackage[mathscr]{euscript}
\usepackage{amsmath, amssymb}
\newcommand{\p}{\partial}
\newcommand{\fb}{\bar{\varphi{}}}
\newcommand{\Rt}{\tilde{R}}
\newcommand{\at}{\tilde{a}}
\newcommand{\te}{\tilde{t}}
\newcommand{\erfi}{\text{erfi}}

\newcommand{\e}{\text{e}}

\title{\boldmath $f(R)$ Dual Theories of Quintessence : Expansion-Collapse Duality}

\author{Dipayan Mukherjee,}
\author{H. K. Jassal}
\author{and Kinjalk Lochan}
\affiliation{Indian Institute of Science Education and Research Mohali,\\ SAS Nagar, Mohali-140306, Punjab, India}

\emailAdd{ph17011@iisermohali.ac.in}
\emailAdd{hkjassal@iisermohali.ac.in}
\emailAdd{kinjalk@iisermohali.ac.in}

\abstract{The accelerated expansion of the universe demands presence of an exotic matter, namely the
  dark energy. Though the cosmological constant fits this role very well, a scalar field minimally
  coupled to gravity, or quintessence, can also be considered as a viable alternative for the
  cosmological constant. We study $f(R)$ gravity models which can lead to an effective description
  of dark energy implemented by quintessence fields in Einstein gravity, using the Einstein
  frame-Jordan frame duality. For a family of viable quintessence models, the reconstruction of
  the $f(R)$ function in the Jordan frame consists of two parts. We first obtain a perturbative
  solution of $f(R)$ in the Jordan frame, applicable near the present epoch. Second, we obtain an
  asymptotic solution for $f(R)$, consistent with the late time limit of the Einstein frame if the
  quintessence field drives the universe. We show that for certain class of viable quintessence
  models, the Jordan frame universe grows to a maximum finite size, after which it begins to
  collapse back. Thus, there is a possibility that in the late time limit where the Einstein frame
  universe continues to expand, the Jordan frame universe collapses. The condition for this
  expansion-collapse duality is then generalized to time varying equations of state models, taking
  into account the presence of non-relativistic matter or any other component in the Einstein frame
  universe. This mapping between an expanding geometry and a collapsing geometry at the field
  equation level may have interesting potential implications on the growth of perturbations therein
  at late times.}

\begin{document}
\maketitle
\flushbottom

\section{Introduction}
\label{sec:introduction}


Observational evidence shows that currently the universe is going through a phase of accelerated
expansion~\cite{riess1998,Perlmutter_1999}. This observation necessitates the existence of an exotic
fluid that violates the `Strong Energy Condition'~\cite{Curiel2014,padmanabhan2010,carroll2019},
referred to as the dark energy (DE)~\cite{amendola2010,copeland2006,wang2010}. The cosmological
constant ($\Lambda$) model is the simplest implementation of dark energy in Einstein's general
theory of relativity~\cite{Padmanabhan_2003,carroll2001,copeland2006}. The $\Lambda$CDM ($\Lambda$
plus Cold Dark Matter) model is also consistent with
observations~\cite{turner84,Efstathiou1990,fujii91,kofman93,Ostriker1995,Betoule_2014}, it is in
fact the simplest and most widely used model to describe dark energy. The energy density contributed
by $\Lambda$ is conventionally associated with the energy density of vacuum.  From a theoretical
point of view, the Lagrangian for the cosmological constant model, i.e., the Einstein-Hilbert action
with an added constant ($\mathcal{L} \sim \sqrt{-g}(R - 2 \Lambda)$), has somewhat a unique status,
as in $1+3$ dimensional spacetime it is arguably the simplest generally covariant Lagrangian, that
leads to second order equations of motion~\cite{lanczos38,lovelock72}. The energy density
corresponding to $\Lambda$ does not vary with the scale factor of the universe and the equation of
state parameter ($w=P/\rho$, where $P,\rho$ are pressure and energy density respectively) associated
with $\Lambda$ is precisely a constant, $w_\Lambda = -1$. Although $w_{\text{DE}} = -1$ is
consistent with observations, deviation from this value is not completely ruled out. Apart from
this, the cosmological constant model suffers from \emph{the fine tuning problem}
~\cite{carroll2001,copeland2006,Padmanabhan_2003}. For example, a naive estimate of the vacuum
energy density of quantum fields, integrated up to the Planck length cut off, can be given as
$\rho_{\text{vacuum}} \sim \hbar k_{\text{Planck}}^4 \sim 10^{74} \text{GeV}^4$, whereas,
observations predict the energy density of the cosmological constant to be
$\rho_{\Lambda} \sim 10^{-47} \text{GeV}^4$, which has a huge discrepancy of the order of $121$ with
the theoretical estimate. This discrepancy can only be rectified by fine tuning
$\rho_{\text{vacuum}}$, which is one of the shortcomings of the $\Lambda$CDM
model~\cite{carroll2001,copeland2006,Padmanabhan_2003}.  It has also been pointed out that a higher
value of the $\Lambda$ would make dark energy take over the matter energy density
($\rho_{\text{m}}$) at an early epoch. If this would have happened at a sufficiently early time, the
accelerating expansion could have prevented the possibility of structure formations in the
universe. The cosmological constant is needed to be fine tuned even in the very early universe, such
that the dark energy-matter equality ($\rho_{\Lambda}/\rho_{\text{m}} \sim \mathcal{O}(1)$) occurs
during the current epoch.

To overcome these problems with the cosmological constant, several dynamical models of dark energy
have been proposed. Unlike the cosmological constant model, a time-dependent model of dark energy
can be fitted with the observations from the current accelerating phase, where it really makes the
impact.  A scalar field minimally coupled to gravity, or quintessence, was introduced to provide a
simple dynamical description of dark energy (see, for example,~\cite{Dolgov82, Weiss1987,ratra88,
  fujii90, Caldwell_1998, peebles88, wetterich88, Ferreira98, brax00, Barreiro00, Zlatev99,
  albrecht00, Nojiri06a}).  The potential ($V$) of the quintessence field ($\varphi$) can be set up
to be sufficiently flat during the current epoch, such that the kinetic term of the field becomes
negligible ($\p_\mu \varphi \sim 0$). This makes the equation of state parameter of the field,
$w_{\varphi} \to -1$, hence the quintessence field mimics the cosmological constant model at the
current epoch (see~\cite{amendola2010,copeland2006,wang2010} and their references for detailed
review).


There is also the possibility that an explanation for dark energy might arise from the outside of
Einstein's gravity framework. A simple form of such extended gravity theories is $f(R)$ gravity,
where instead of Ricci scalar $R$, a general function of the curvature scalar, $f(R)$, is used in
the Einstein-Hilbert action (see, for example,~\cite{Elizalde:2018rmz, Odintsov:2020qzd,
  Astashenok:2020isy, casadio2019,vasilev2019,barrow2019, choudhury2019, hough2019,
  Elizalde:2018rmz, Odintsov20, Oikonomou21} and references therein for recent works in $f(R)$ gravity, for
reviews see~\cite{defelice2010,sotiriou2010,faraoni2014}). There are roughly two approaches one can
take to describe dark energy vis-a-vis $f(R)$ gravity.


The first approach is to treat modified gravity as a `correction' to Einstein's gravity, such that
the `correction' itself is responsible for the acceleration of the universe. Here it is convenient
to treat the deviation of the Einstein field equation from the modified field equation as an
effective energy momentum tensor of a perfect fluid. One of the early examples of this approach can
be found in~\cite{starobinsky1980} in the context of inflation. Later it was extensively studied in
the context of late-time acceleration as well (for example, see~\cite{capozziello2003, dolgov2003,
  Nojiri06, Nojiri09}). The cosmological viability of $f(R)$ in such models was discussed
in~\cite{amendola2007,defelice2010,sotiriou2010}.

There is also another approach where the dark energy implemented by a quintessence field (in
Einstein gravity) can effectively be studied as a pure gravity theory governed by an $f(R)$ action.
It is well known that a conformal transformation of the metric of the $f(R)$ action can lead to a
theory of Einstein's gravity with a quintessence field, in a conformally connected spacetime (see,
for example,~\cite{maeda1989, wands1994, faraoni1998, capozziello06, Briscese07, Elizalde08,
  wald2009, sotiriou2010, defelice2010, Nojiri_2011, Capozziello_2011, Qiu12, Bahamonde16,
  Bahamonde17, Nashed20}). The description of the universe, where the gravity action becomes
Einstein-Hilbert action, is referred to as the `Einstein frame', whereas, the initial description is
referred to as the `Jordan frame'.


The conformal parameter of this transformation is given by the $f(R)$ model itself. This establishes
a duality between $f(R)$ gravity without quintessence field in Jordan frame and Einstein gravity
with quintessence field in Einstein frame, such that, an $f(R)$ function corresponds to a
quintessence potential. Due to this duality, one can treat a quintessence model in Einstein gravity
as an $f(R)$ gravity model, without the need of a quintessence field.

In this paper, we are interested mainly in quintessence models with an equation of
state $w(a) = w_0 - w' \ln a$.  The corresponding quintessence potential has a closed and relatively
simple analytical form. This parameterization was first introduced in~\cite{Efstathiou_1999} to fit
with three tracking quintessence models. It was shown to be a good fit with observations in the
redshift range $z \lesssim 4$. This model was further extended in~\cite{Feng_2011} to account for
observation from a wider range of redshift parameter. Models with logarithmic $w(a)$ were further
constrained from the SNIa+BAO+H($z$) data in~\cite{tripathi2017,sangwan2018}. We obtain a class
of $f(R)$ theories that can recover these potentials in Einstein frame.  The reconstruction of
the $f(R)$ function can be broken in to two parts. For the near current time in the Einstein frame,
we find perturbative solutions of $f(R)$ which are valid in the small Jordan frame curvature limit.
From this we estimate the perturbative solution which is most suited in the current epoch of the
Einstein frame universe. For the distant future in Einstein frame, we obtain an asymptotic solution
for $f(R)$.

We further show that the logarithmic parameterization of $w(a)$ belongs to a class of quintessence
models for which the Jordan frame scale factor has a finite maximum value. In the late time limit of
the Einstein frame universe, the Einstein frame scale factor increases indefinitely, while the
Jordan frame universe collapses after attaining a maximum. A general condition for the collapse of
the Jordan frame is then obtained which takes into account other components of the universe. Using
this we find that the presence of dust prevents the Jordan frame collapse. Finally, we show that the
introduction of positive spatial curvature~\cite{Valentino_2019} may still allow the
expansion-collapse duality in the presence of non-relativistic matter.

The general condition for the expansion-collapse duality of the Einstein and Jordan frame can be
used in further studies to explore other viable quintessence models. An expanding universe with
quintessence field can also be looked at as a collapsing universe with different equations of
motion. Such a correspondence between expanding and collapsing geometries can have applications in
studies of growth of cosmological perturbations. For example, one can study the back reaction of the
curvature and matter perturbations in our universe collectively as a gravitational perturbation of a
collapsing geometry leading to a good estimate of back reaction at late times.

The paper is organized as follows. In Sec.~\ref{sec:reconstr-dark-energy} we review the
quintessence model and the reconstruction of the quintessence field potential. In
Sec.~\ref{sec:reconstr-fr-models} we obtain the analogous $f(R)$ theories consistent with the
quintessence model; first perturbatively near current era and then for remote future. A relation
between the scale factors of the FRW universes in the Einstein and the Jordan frame is then obtained
in Sec.~\ref{sec:comparison-late-time}. Here we demonstrate that at late times the Jordan frame
universe collapses back unlike the universe driven by the quintessence field in the Einstein frame.
In Sec.~\ref{sec:expans-coll-dual} we derive a general condition for the expansion-collapse duality
and further explore the effect of dust and spatial curvature on the collapse of the Jordan frame. We
conclude the paper with a summary and discussion of implications in Sec.~\ref{sec:concl-disc}.

\section{Reconstruction of quintessence field potentials}
\label{sec:reconstr-dark-energy}
The accelerated expansion of the universe requires the presence of dark energy, characterized by an
equation of state parameter $w < -1/3$, this is generally termed as violation of the `strong energy
condition'. A scalar field minimally coupled to gravity, or quintessence, is a viable candidate for
dark energy (see~\cite{copeland2006,tsujikawa2013,amendola2010}).  In this section we briefly review
the quintessence models corresponding to constant and logarithmic equation of state parameters.

\subsection{Quintessence model of dark energy}
\label{sec:quint-model-dark}

The action of a generic quintessence field in Einstein gravity is give by
\begin{align}
  \label{eq:1}
  \mathcal{S}_Q = \int d^4x \sqrt{-g} \frac{R}{2 \kappa^2} - \int d^4x \sqrt{-g} \left( \frac{1}{2}\partial _\mu\Phi \partial ^\mu \Phi +V(\Phi) \right).
\end{align}
Here $\kappa^2 = 8 \pi G$, $R$ is Ricci scalar, and $g_{\mu \nu}$ is taken to be the spatially flat FRW metric,
\begin{align*}
  g_{\mu \nu} \dot{=} \text{diag}\left(-1, a^2(t), a^2(t), a^2(t) \right).
\end{align*}
The quintessence field $\Phi$ is taken as a function of time only. The definition of the
energy-momentum tensor,

\begin{align}
  \label{eq:2}
  T_{\mu \nu} = \frac{-2}{\sqrt{-g}}\frac{\delta S_\Phi}{\delta g^{\mu \nu}}
\end{align}
leads to energy density ($\rho$) and pressure ($P$) associated with the field,
\begin{align}
  T_{00} &= \rho = \frac{\dot{\Phi}^2}{2}+V \label{eq:3}\\
  T^i_i &= P = \frac{\dot{\Phi}^2}{2}-V  \label{eq:4},
\end{align}
where the dot represents derivative with respect to comoving time. From this one can recover the
quintessence potential and the time derivative of the field as,
\begin{align}
  \label{eq:5}
  V &= \frac{1}{2}(1 - w(a))\rho(a),
\end{align}
\begin{equation}
  \label{eq:6}
  \dot{\Phi}^2 = (1 + w(a))\rho(a),
\end{equation}
leading to the equation of state parameter $w(a)$,
\begin{align}
  \label{eq:7}
  w(a) = \frac{P}{\rho} = \frac{\frac{1}{2} \dot{\Phi}^2 - V(\Phi)}{\frac{1}{2} \dot{\Phi}^2 + V(\Phi)}.
\end{align}
This shows that for a slowly rolling quintessence field, i.e., if the kinetic term of the field
becomes negligible with respect to the potential ($V(\Phi) \gg \dot{\Phi}^2 $), then $w \to -1$,
therefore the quintessence field becomes capable of driving the acceleration.

For example, if the equation of state parameter ($w$) is time-independent, then the continuity
equation
\begin{subequations}
  \label{eq:8}
  \begin{align}
    \label{eq:9}
    \nabla_\mu T^{\mu \nu} &= 0,\\
    \label{eq:10}
    \dot{\rho} + 3 H \rho (1 + w) &= 0
  \end{align}
\end{subequations}
can be solved to obtain the energy density as $\rho(a) = \rho_0 a^{-3(1 + w)}$, where
$H = \dot{a}/a$ is the Hubble parameter. Using this in~\eqref{eq:6}, together with the Friedmann
equation
\begin{equation}
  \label{eq:11}
  H^2 = \frac{\kappa^2}{3}\rho(a),
\end{equation}
one can obtain the scale factor dependence of the field $\Phi$ as
\begin{align}
  \label{eq:12}
  \varphi(a) = \Phi(a) - \Phi_0 &= \sqrt{\frac{3(1+w)}{\kappa^2}} \ln {a},
\end{align}
here $\Phi_0$ is the value of the field at current scale factor $a = 1$.
In this case the reconstructed quintessence potential becomes~\cite{sangwan2018}
\begin{align}
  \label{eq:13}
  V(\varphi) = \frac{1}{2}(1-w)\rho_0\exp \left( -\sqrt{3\kappa^2 (1+w)}(\varphi)\right).
\end{align}

\subsection{Quintessence with logarithmic equation of state parameter}
\label{sec:quint-with-logar}

The cosmological constant model provides a constant energy density of dark energy with
$w_{\text{DE}} = -1$. In general, $w$ can take other values and can also vary with
time. Time-dependence in $w$ is often introduced by postulating a functional form of $w(a)$ with
adjustable parameters. These parameters of the model $w(a)$ can then be constrained from different
cosmological observations. Several parameterizations of $w(a)$ have been
proposed~\cite{Efstathiou_1999, chevallier01, Linder_2003, Feng_2011, Jassal_2005} (for reviews
see~\cite{Bamba_2012, Capozziello_2006, copeland2006, tripathi2017, colgain21}).

Given a parameterization of $w(a)$, one can, in principle, reconstruct the corresponding
quintessence field potential $V(\Phi)$. In this paper we reconstruct the $f(R)$ action corresponding
to the quintessence model, where the functional form of $V(\Phi)$ plays a crucial part. We consider
the logarithmic parameterization of $w(a)$, which leads to a quintessence potential with a closed
and relatively simple functional form; the parameterization is given by
\begin{equation}
  \label{eq:14}
  w(a) = w_0 - w' \ln a,
\end{equation}
where $w_0$ and $w'$ are the constant parameters. This parameterization was first introduced
in~\cite{Efstathiou_1999} to fit several tracking quintessence models, and it was shown to be
consistent with low redshift observations. Later on a modification on this parameterization was
proposed in~\cite{Feng_2011} to make this model compatible in the $z \to \infty$ limit. Recently,
this parameterization was constrained from SNIa+BAO+H($z$) data in~\cite{tripathi2017} and the
corresponding quintessence potential was reconstructed in~\cite{sangwan2018}. The ranges of the
quintessence parameters within $3\sigma$ confidence level, compiled from SNIa+BAO+H($z$) data is
provided in~\cite{tripathi2017,sangwan2018} as
\begin{align}
  \label{eq:15}
  -1.09 \le w_0 \le -0.66,\ -1.21 \le w' \le 0.25,\ 0.26 \le \Omega_M \le 0.32.
\end{align}

One can solve the continuity equation~\eqref{eq:10} for this model to obtain the energy density in
terms of the scale factor as
\begin{align}
  \label{eq:16}
  \rho(a) &= \rho_0 a^{-3(1+w_0)+ \frac{3}{2} w' \ln a},
\end{align}
where $\rho_0 = \rho(a=1)$. In the limit of $a \to \infty$, the energy density becomes $0$ if
$w'<0$, and $\infty$ for $w'>0$. The quintessence field as a function of the scale factor can be
obtained as
\begin{align}
  \label{eq:17}
  \varphi(a) = \Phi(a) - \Phi_0 &= \frac{2}{\sqrt{3}} \frac{1}{\kappa w'} \left[ (1+w_0)^{\frac{3}{2}} - {(1+w_0-w' \ln a)}^ {\frac{3}{2}}\right].
\end{align}
We note that for $w'>0$, the field becomes complex in the limit of $a \to \infty$. In order for
$\varphi$ to be real in between the current scale factor ($a_0 = 1$) and $a \to \infty$, we will
exclusively consider the model with $w'<0$. The potential of the quintessence field can be
reconstructed as
\begin{align}
  \label{eq:18}
  V(\varphi) &= \rho_0 \frac{1}{2}\left[ 2-\left\{
               {(1+w_0)}^{\frac{3}{2}}-\frac{3w'}{2 \sqrt{3}} \kappa
               (\varphi) \right\}^{\frac{2}{3}}\right] \nonumber\\
             &\exp
               \left(-\frac{3}{2w'}\left\{
               (1+w_0)^2-\left[(1+w_0)^{\frac{3}{2}}-\frac{3w'}{2\sqrt{3}}\kappa(\varphi)\right]^{\frac{4}{3}}\right\}
               \right).
\end{align}
See~\cite{sangwan2018} for detailed derivations of the quintessence potentials for both
time-dependent and independent equation of state parameters.

\section{Reconstruction of $f(R)$ models}
\label{sec:reconstr-fr-models}

In this section we reconstruct $f(R)$ functions in the Jordan frame which map to quintessence models
discussed above, in the Einstein frame.

\subsection{The Einstein frame-Jordan frame duality}
\label{sec:conf-transf-einst}

$f(R)$ gravity theory is a simple extension of Einstein gravity, where the Einstein-Hilbert action
is modified by replacing the Ricci scalar ($R$) with a general function of Ricci scalar $f(R)$
(see~\cite{defelice2010,sotiriou2010,Capozziello_2011,Nojiri17} for detailed review). The action of $f(R)$
theory is given by
\begin{align}
  \label{eq:19}
  \mathcal{S}_J &= \frac{1}{2\kappa^2} \int d^4 x \sqrt{-g}f(R).
\end{align}
The description of the universe obtained from this action is referred to as the `Jordan frame'.  A
duality between quintessence field in Einstein gravity and $f(R)$ gravity can be established
through a conformal transformation on the metric of the $f(R)$ action. The conformal transformation
\begin{align}
  \label{eq:20}
  \tilde{g} _{\mu \nu} = \Omega^2(x) g _{\mu \nu},
\end{align}
with the choice of the conformal parameter $\Omega^2 = \p_R f(R) = F(R)$, allows for the
action~\eqref{eq:19} to be written as
\begin{align}
  \label{eq:21}
  \mathcal{S}_E = \int d^4x \sqrt{-\tilde{g}} \frac{1}{2\kappa^2} \tilde{R}-\int d^4x \sqrt{-\tilde{g}} \left[ \frac{1}{2} \tilde{g}^{\mu \nu} \partial_\mu \varphi \partial_\nu \varphi + V(\varphi)\right],
\end{align}
where $\Rt$ is the Ricci scalar corresponding to the metric $\tilde{g}_{\mu \nu}$, the scalar
field $\varphi$ and the potential $V$ are identified as
\begin{align}
  \label{eq:22}
  \varphi =  \frac{\sqrt{6}}{2\kappa} \ln F 
\end{align}
and
\begin{align}
  \label{eq:23}
  V = \frac{1}{2\kappa^2}\frac{FR-f}{F^2}
\end{align}
(see, for example,~\cite{wald2009,defelice2010,sotiriou2010} for more details). Universe described
using the action~\eqref{eq:21} is referred to as the `Einstein frame'. The metric in the Einstein
frame is taken to be a spatially flat FRW metric,
i.e., $\tilde{g} _{\mu \nu} \dot{=} \text{diag} (-1, \tilde{a}^2, \tilde{a}^2, \tilde{a}^2)$,
where $\tilde{a}$ is the scale factor in Einstein frame.

The Einstein frame action~\eqref{eq:21} represents a quintessence field in Einstein gravity. The
definition of the quintessence potential~\eqref{eq:23} is key to the duality between Einstein and
Jordan frames, i.e., a quintessence potential corresponds to a $f(R)$ function in Jordan frame
(see~\cite{nojiri2019,Nojiri_2011,KANEDA_2010,Schmidt1997,Ketov_2015,Capozziello_2006b,Hu_2007,Wetterich_2014},
for detailed review see~\cite{faraoni2014,defelice2010,sotiriou2010}). Now we will use this
equivalence to constrain $f(R)$ for the quintessence potentials discussed in
Sec.~\ref{sec:reconstr-dark-energy}.

\subsection{Time-independent equation of state parameter}
\label{sec:time-independent-eos}

We start with deriving the function $f(R)$ such that the Jordan frame becomes dual to a quintessence
field with constant equation of state parameter in Einstein frame. We identify the
field term in~\eqref{eq:12} as $\Phi-\Phi_0 = \varphi =  \sqrt{6} \ln F/(2 \kappa)$. The
potential~\eqref{eq:13} becomes
\begin{align}
  \label{eq:24}
  V &=\frac{1}{2} (1-w) \rho_0 F^{-\frac{3}{\sqrt{2}}\sqrt{1+w}}.
\end{align}
In order for this potential to be generated by the $f(R)$ function,~\eqref{eq:24} should be the same as
$V$ in~\eqref{eq:23}, i.e., $f(R)$ should satisfy the following differential equation
\begin{subequations}
  \label{eq:25}
  \begin{align}
    \frac{1}{2\kappa^2}\frac{FR-f}{F^2} &= \frac{1}{2}(1-w)\rho{}_0 F^{-\frac{3}{\sqrt{2}}\sqrt{1+w}},\\
    A_1F^b -FR + f &= 0,
  \end{align}
\end{subequations}
where,
\begin{subequations}
  \begin{align}
    \label{eq:26}
    A_1 &= \kappa^2 \rho{}_0 (1-w) \\
    \label{eq:27}
    b &= -\frac{3}{\sqrt{2}}\sqrt{1+w}+2
  \end{align}
\end{subequations}
Equation~\eqref{eq:25} can be solved analytically and it has a trivial linear solution
\begin{align}
  \label{eq:28}
  f(R) = C_1R - A_1{C}_1^b
\end{align}
where $C_1$ is an integration constant. The conformal parameter in this case becomes a constant,
$C_1$, thus we recover the cosmological constant model.
The non-trivial solution of~\eqref{eq:25} has a power law form,
\begin{align}
  \label{eq:29}
  f(R) = \frac{b-1}{b{(A_1 b)}^{\frac{1}{b-1}}} R^{\frac{b}{b-1}}
\end{align}
where $A_1, b \neq 0$. Interestingly, this $f(R)$ model never reduces to the cosmological constant
model for any finite value of the parameter $b$.

\subsection{Logarithmic equation of state parameter}
\label{sec:logarithmic-eos}

Now we reconstruct $f(R)$ models which lead to a scalar field potential consistent with the
logarithmic equation of state parameter~\eqref{eq:14}. In this case, it is convenient to start with
a parametric solution of the $f(R)$ function.

Since $\ln F = 2 \kappa \varphi/ \sqrt{6}$, we have 
\begin{align}
  \label{eq:30}
  \diff{V}{\varphi} &=   \frac{1}{\sqrt{6} \kappa} \frac{2f - FR}{F^2}.
\end{align}
Adding~\eqref{eq:30} with~\eqref{eq:23} and rearranging the terms we get
\begin{align}
  \label{eq:31}
  f(\fb) &=  \exp \left( 2 \fb \right)\left\{ 2 \kappa^2 V + 2 \kappa^2 \diff{V}{\fb}\right\},
\end{align}
where we have re-scaled the quintessence field to be dimensionless as
$\fb = \ln F = 2 \kappa \varphi/\sqrt{6}$. Similarly, subtracting~\eqref{eq:30} from~\eqref{eq:23} and
rearranging we get
\begin{align}
  \label{eq:32}
  R(\fb) &= \exp \left( \fb \right) \left\{ 4 \kappa^2 V + 2 \kappa^2 \diff{V}{\fb} \right\}.
\end{align}
Equations~\eqref{eq:31} and~\eqref{eq:32} represent a parametric solution for the $f(R)$ function
consistent with an arbitrary quintessence potential $V(\varphi)$, where the field $\fb$ plays
the role of the parameter. For the potential~\eqref{eq:18}, equations~\eqref{eq:31}
and~\eqref{eq:32} become
\begin{align}
  \label{eq:33}
  f(\fb) = C \exp\left( 2 \fb + C_2 \mathcal{P}^{\frac{4}{3}} \right) \left[ 2 - \mathcal{P}^{\frac{2}{3}} + \frac{2}{3} B \mathcal{P}^{-\frac{1}{3}} - \frac{8}{3} B C_2 \mathcal{P}^{\frac{1}{3}} + \frac{4}{3} B C_2 \mathcal{P}  \right]
\end{align}
and
\begin{align}
  \label{eq:34}
  R(\fb) = C \exp \left(\fb + C_2{\mathcal{P}}^{\frac{4}{3}}\right)\left[ 4 - 2{\mathcal{P}}^{\frac{2}{3}} + \frac{2}{3}B {\mathcal{P}}^{-\frac{1}{3}}-\frac{8}{3} B C_2 {\mathcal{P}}^{\frac{1}{3}}+ \frac{4}{3} B C_2 \mathcal{P}\right],
\end{align}
where we have defined the following constants
\begin{align}
  \label{eq:35}
  {(1+w_0)}^2=C_1,\ \frac{3}{2w'}=C_2,\ {(1+w_0)}^{\frac{3}{2}}=A,\ \frac{3 \sqrt{2}}{4} w'=B,\ C' = \rho_0e^{-C_1 C_2}, C = \kappa^2 C'
\end{align}
and $\mathcal{P} = A - B \ln F = A -  B\fb$.

\subsubsection{Perturbative solution for $F(R)$}
\label{sec:pert-solut-fr}
We use an analytical approximation to invert equation~\eqref{eq:34} to obtain $F(R)$. Specifically,
we consider the $f(R)$ theory to be a small deviation from the Einstein gravity, such that, the
limit $R \to 0$ recovers the Einstein gravity. We assume that the $f(R)$ function has a form
\begin{align}
  \label{eq:36}
  f(R) = R + \epsilon \psi(R),
\end{align}
$\epsilon$ is a constant parameter such that $|\epsilon \psi(R)/R|\ll 1$. As $R \to 0$, $f(R)$
theories are often subjected to the following two stability criteria
\cite{defelice2010,sotiriou2010,dolgov2003,Faraoni_2006}
\begin{subequations}
  \label{eq:37}
  \begin{align}
    f' &= F > 0\\
    f'' &= F' >0.
  \end{align}
\end{subequations}
In an effective fluid description of $f(R)$ gravity, the contribution from the $f(R)$ action in the
field equation is treated as an energy-momentum tensor of an effective fluid (reviews can be found
in~\cite{defelice2010,sotiriou2010}). Within this description, the term $\kappa^2/F$ appears as the
effective gravitational coupling term. Thus, the first condition $F>0$ is required to ensure an
overall positive gravitational coupling. On the other hand, when the $f(R)$ model is taken to be a
dual description of a quintessence model, as in the current study, $F>0$ is necessary for the
quintessence field to be real (see~\eqref{eq:22}). The second condition $F'>0$ is to avoid the
Dolgov-Kawasaki or matter instability~\cite{Faraoni_2006,dolgov2003}. Taking into account the form
of $f(R)$ from~\eqref{eq:36} and considering weak gravity regime, the metric and Ricci scalar can be
perturbed as
\begin{subequations}
  \label{eq:38}
  \begin{align}
    \label{eq:39}
    g _{\mu \nu} &= \eta _{\mu \nu} + h _{\mu \nu},\\
    \label{eq:40}
    R &= 0 + \delta R,
  \end{align}
\end{subequations}
where $\eta_{\mu \nu}$ is the Minkowski metric, Ricci scalar is perturbed over the background value
$0$. It can be shown that the field equation of $\delta R$ (up to first order in $\delta R$) has an
effective mass-square term $\sim ( 3 \varepsilon \psi'' )^{-1}$, the sign of which is determined by
$F'$ using~\eqref{eq:36} (see~\cite{sotiriou2010,Faraoni_2006} for the derivation). Hence, it is
argued that the Ricci scalar perturbation $\delta R$ is stable only for $\psi''>0$ or $F'>0$.

Before using the ansatz~\eqref{eq:36} in the current model, we first check how such perturbative
solution can lead to a relation between the perturbation in the quintessence field in the Einstein
frame and Ricci scalar perturbation in the Jordan frame. By considering a matter-less Jordan frame
universe in the weak gravity limit, as in~\eqref{eq:38}, and using the Einstein-Jordan frame
correspondence, $2 \kappa \varphi = \sqrt{6} \ln F$, one can obtain the perturbation in the
quintessence field as
\begin{align}
  \label{eq:41}
  \kappa \delta \varphi  = \frac{\sqrt{6}}{2} \left. \frac{F'}{F} \right|_{R=0} \delta R = \frac{\sqrt{6}}{2} \left. \frac{\epsilon \psi''}{1 + \epsilon \psi'} \right|_{R=0} \delta R,
\end{align}
where the prime denotes derivative with respect to $R$. Under this assumption, the scalar field
perturbation $\delta \varphi$ becomes proportional to the Ricci perturbation in the Jordan
frame, $\delta R$. This relation may reveal interesting duality between the different fields in
Einstein and Jordan frames.

With this motivation, we now seek a class of functions $f(R)$ for which $F(R)$ can be expanded in
the powers of $R$, consistent with the current quintessence model. In order to derive an appropriate
analytical form, we truncate the expansion of $F(R)$ after the quadratic term, i.e.,
\begin{equation}
  \label{eq:42}
  F(R) \approx F_A(R) = \epsilon_0 + \epsilon_1 R + \epsilon_2 R^2.
\end{equation}
Here dimensions of the constants $\epsilon_0, \epsilon_1, \epsilon_2$ are such that $F(R)$ is
dimensionless. The above ansatz is valid only within a region of time where $R$ is sufficiently
small, such that the contributions of higher order terms of $R$ in $F(R)$ can be
neglected.

The validity of the ansatz also depends upon the implicit assumption that $F(R)$ is finite valued in
the region where the solution is applicable. The function $F(R)$ has to be consistent with the
Einstein-Jordan frame correspondence, i.e., $F = \exp (2 \kappa \varphi/\sqrt{6})$. It is possible
that at some later time in the Einstein frame the field $\varphi$, and thus $F$, becomes divergent,
however, the Jordan frame curvature $R$ remains finite. This is clearly in conflict
with~\eqref{eq:42}, thus the ansatz becomes invalid in such limits. It is shown later that this is
exactly the case in the late time limit of Einstein frame, in that case we obtain a non-perturbative
asymptotic solution.

We obtain the perturbative solution for $F(R)$ by deriving the coefficients in~\eqref{eq:42},
$\epsilon_0, \epsilon_1, \epsilon_2$, in terms of the parameters of the quintessence models,
$(w_0, w')$. This is done by using the ansatz~\eqref{eq:42} in~\eqref{eq:34} and equating the
coefficients of the powers of $R$ in both sides of the equation, where the terms of the order $R^3$
and higher are neglected. The analytical results and the functions $\epsilon_{0,1,2}(w_0, w')$ are
given in Appendix~\ref{sec:appendix1}. We find that for a given set of $(w_0,w')$ values, there are
four sets of possible $\epsilon_{0,1,2}$, hence there are four possible solutions for the
ansatz~\eqref{eq:42} (see Appendix~\ref{sec:appendix1}).

To illustrate the perturbative solution, let us consider the central values of the ranges of
quintessence parameters provided in~\cite{tripathi2017,sangwan2018},
\begin{align}
  \label{eq:43}
  w_0 = -0.87,\ w' = -0.48.
\end{align}
For these values, the possible real solutions for $\epsilon_{0,1,2}$ are given in the
table~\ref{tab:1}.

\begin{figure}
  \centering
  \begin{subfigure}[b]{0.49\textwidth}
    \centering
    \includegraphics[width=\textwidth]{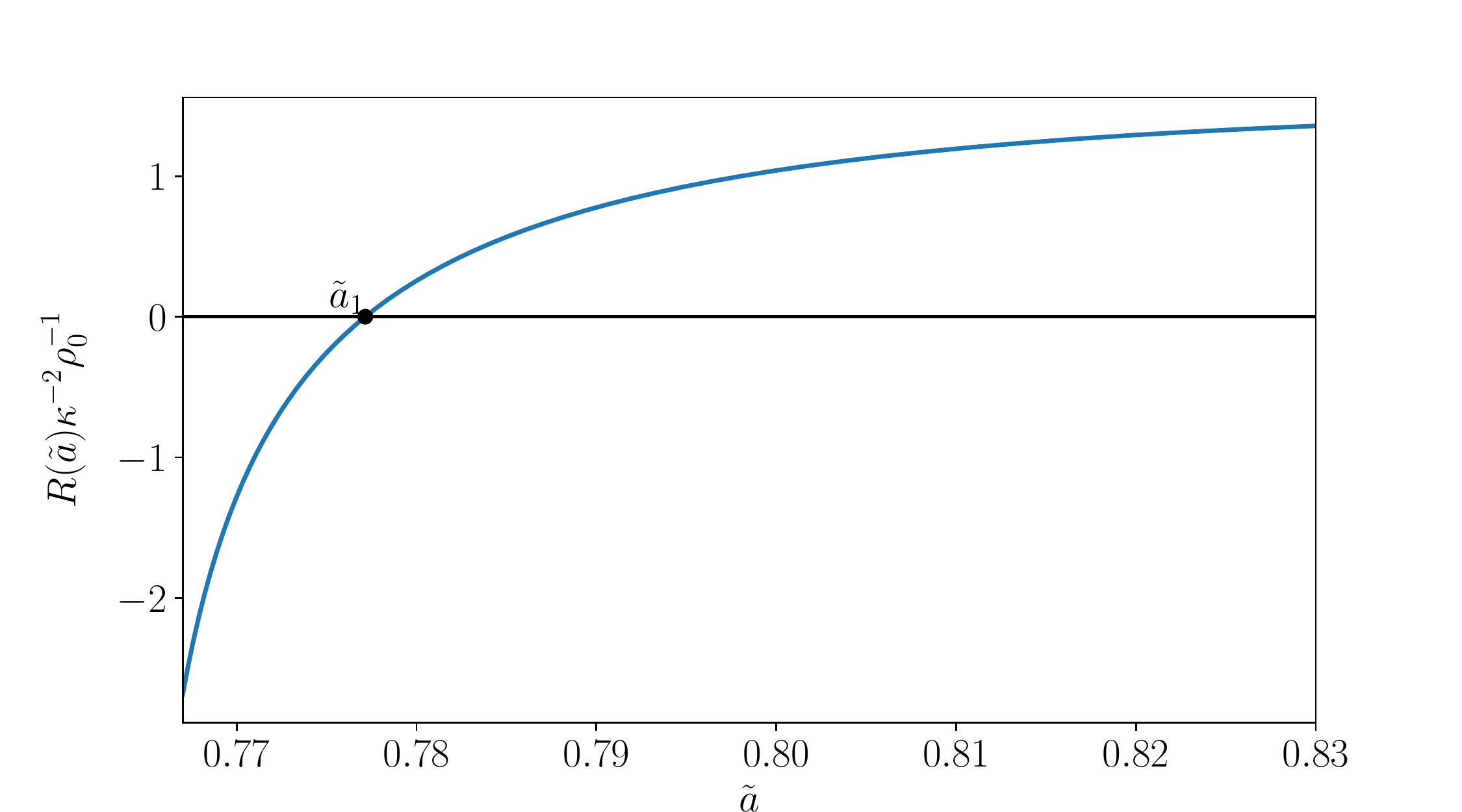}
    \caption{PS I}
    \label{fig:1}
  \end{subfigure}
  \begin{subfigure}[b]{0.49\textwidth}
    \centering
    \includegraphics[width=\textwidth]{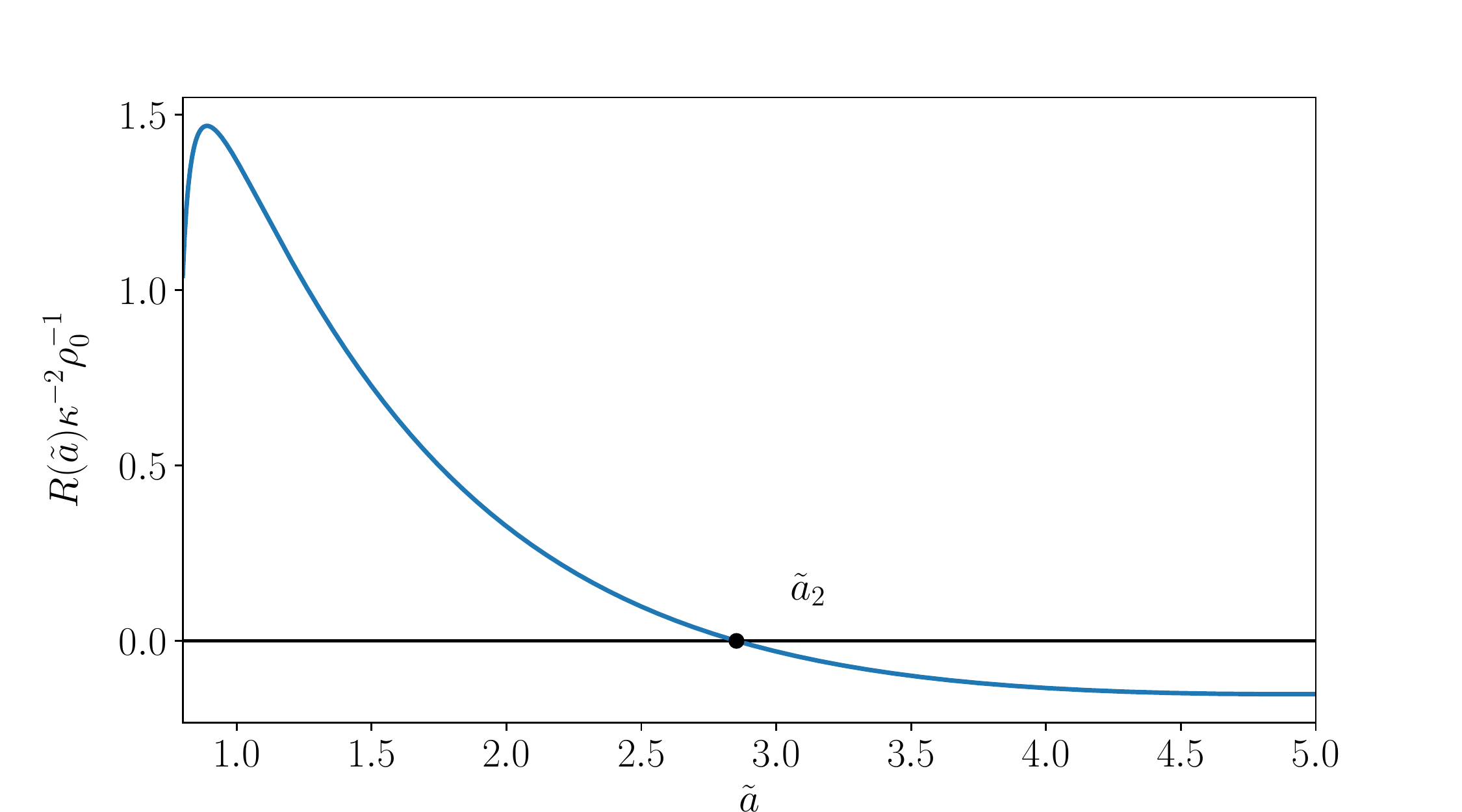}
    \caption{PS II}
    \label{fig:2}
  \end{subfigure}
  \begin{subfigure}[b]{0.49\textwidth}
    \centering
    \includegraphics[width=\textwidth]{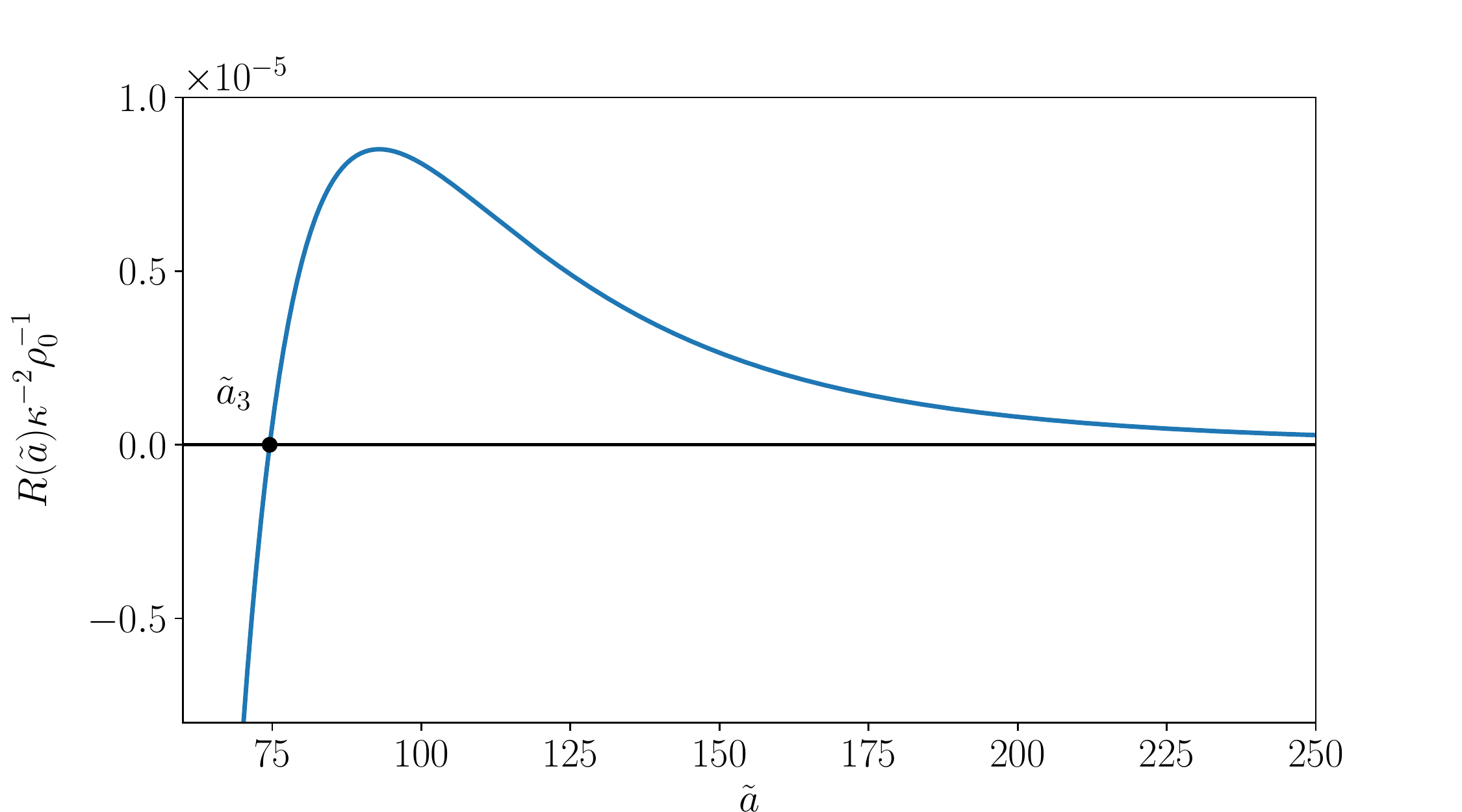}
    \caption{PS III}
    \label{fig:3}
  \end{subfigure}
  \caption{Jordan frame Ricci scalar as a function of Einstein frame scale factor, $R(\tilde{a})$, with
    quintessence parameters $(w_0,w')=(-0.87,-0.48)$. The ansatz~\eqref{eq:42} is applicable in the
    neighborhood of the roots $\tilde{a}_{1,2,3}$. In the limit $\tilde{a} \to 0$, Jordan frame Ricci
    $R \to 0$.}
  \label{fig:4}
\end{figure}

It is useful for further discussion to consider the profile of Jordan frame curvature as a function
of Einstein frame scale factor. The quintessence field ($\varphi$) can be written as a function of
Einstein frame scale factor ($\tilde{a}$) as~\eqref{eq:17}
\begin{align}
  \label{eq:44}
  \varphi(\tilde{a}) = \Phi(\tilde{a}) - \Phi_0 &= \frac{2}{\sqrt{3}} \frac{1}{\kappa w'} \left[ (1+w_0)^{\frac{3}{2}} - {(1+w_0-w' \ln \tilde{a})}^ {\frac{3}{2}}\right].
\end{align}
Using the Einstein-Jordan frame correspondence~\eqref{eq:22}, $F$ is obtained as a function of
$\tilde{a}$ as
\begin{align}
  \label{eq:45}
  F(\tilde{a}) = \exp \left( \frac{2\kappa}{\sqrt{6}}\varphi(\tilde{a})\right) = \exp \left( \frac{2 \sqrt{2}}{3w'} \left[(1 + w_0)^{\frac{3}{2}} - (1 + w_0 - w' \ln \tilde{a})^{\frac{3}{2}} \right] \right).
\end{align}
$F(\tilde{a})$ can be used in~\eqref{eq:34} to obtain $R(\tilde{a})$ (see Fig.~\ref{fig:4}) ($R(\tilde{a})$ has a
complicated expression, it is not shown explicitly). The roots of $R(\tilde{a})$ are identified as
$\tilde{a}_{1,2,3}$ (see table~\ref{tab:1}). $R(\tilde{a})$ is further used in the ansatz $F_A(R)$ to derive $F_A(\tilde{a})$. We
are now in a position to compare the ansatz $F_A(\tilde{a})$ with the exact $F(\tilde{a})$ (from~\eqref{eq:45})
in the neighborhoods of the roots of $R(\tilde{a})$. 

\begin{figure}
  \centering
  \begin{subfigure}[b]{0.49\textwidth}
    \centering
    \includegraphics[width=\textwidth]{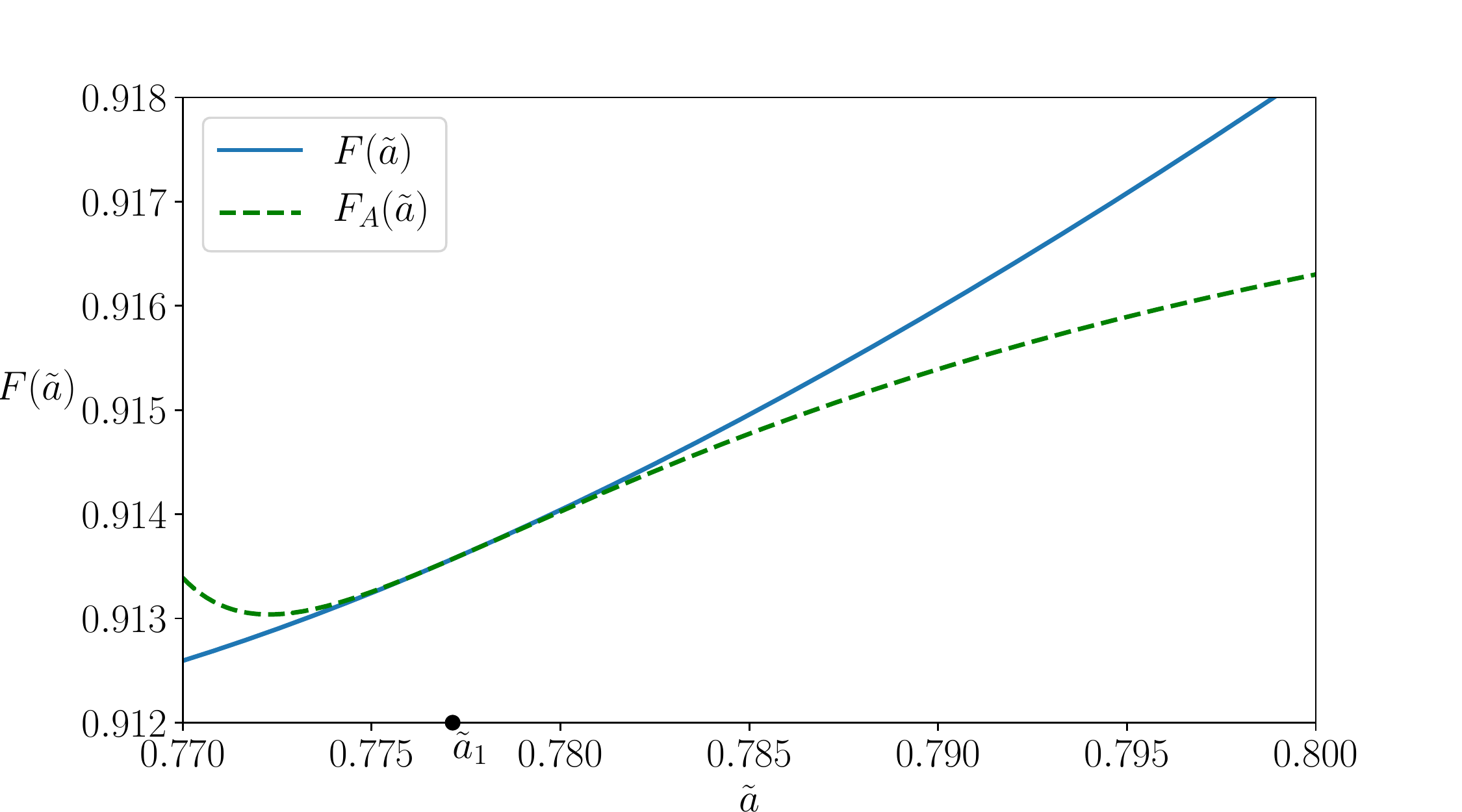}
    \caption{PS I}
    \label{fig:5}
  \end{subfigure}
  \begin{subfigure}[b]{0.49\textwidth}
    \centering
    \includegraphics[width=\textwidth]{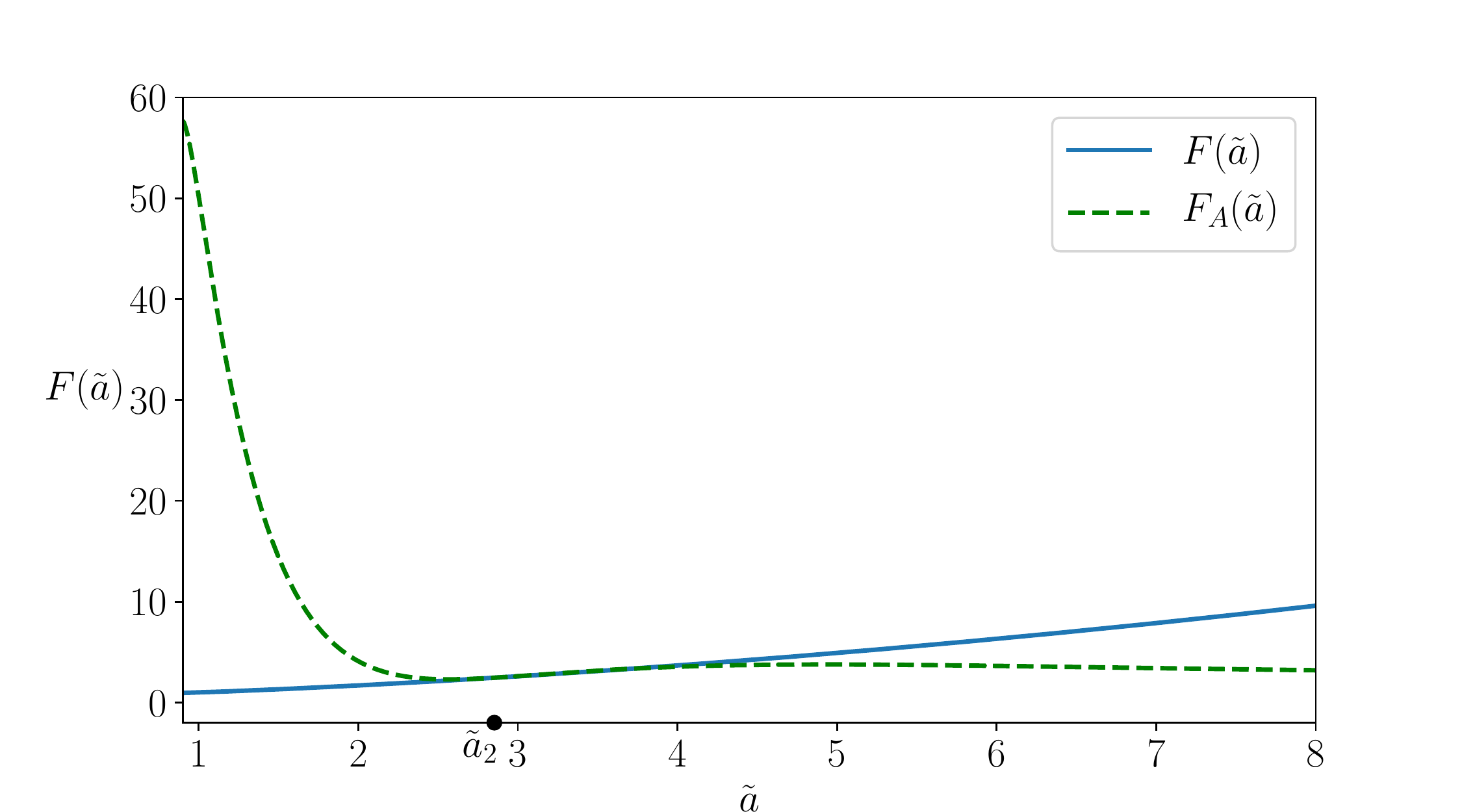}
    \caption{PS II}
    \label{fig:6}
  \end{subfigure}
  \begin{subfigure}[b]{0.49\textwidth}
    \centering
    \includegraphics[width=\textwidth]{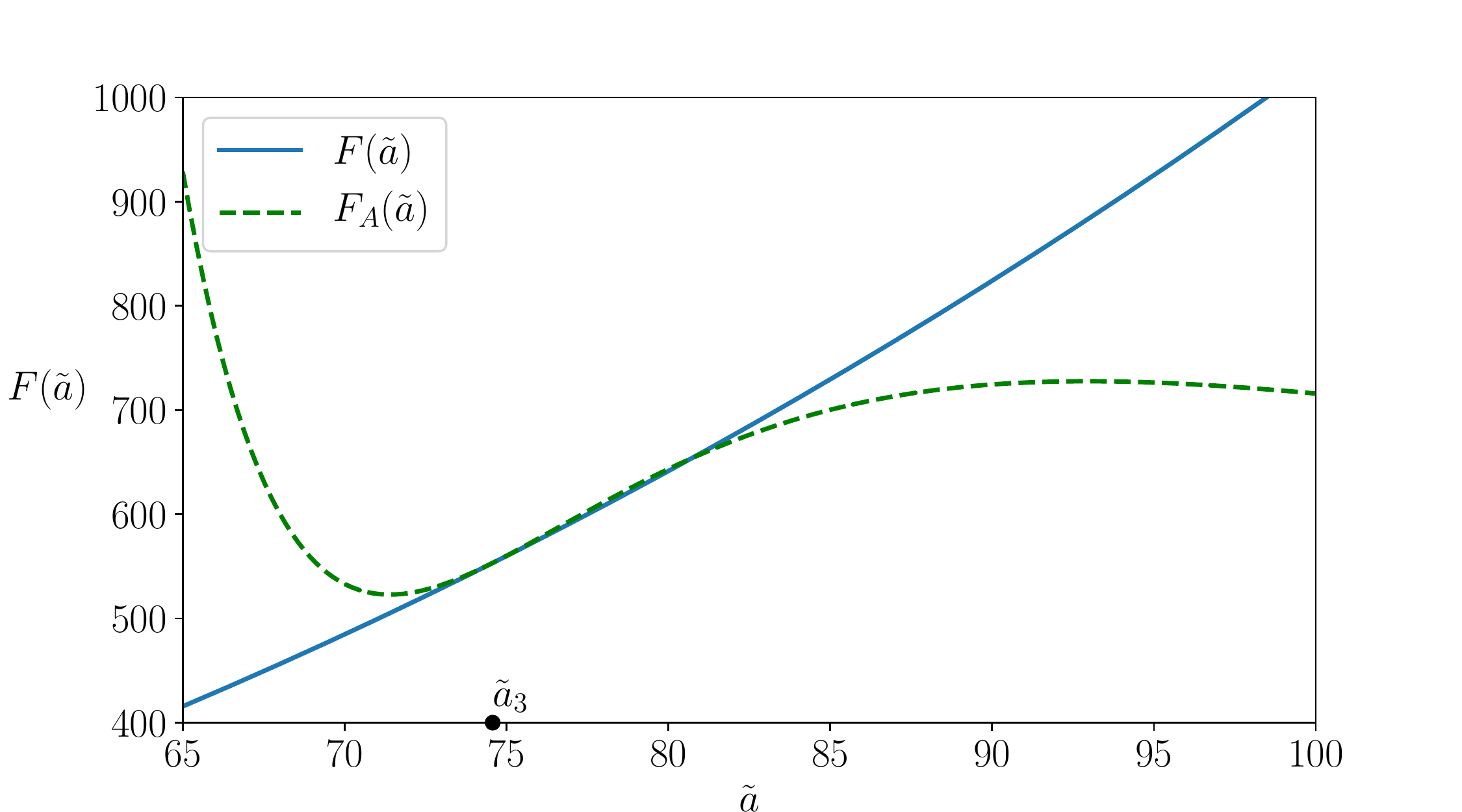}
    \caption{PS III}
    \label{fig:7}
  \end{subfigure}
  \caption{Comparison of $F(\tilde{a})$  and $F_A(\tilde{a})$  for quintessence parameters
    $(w_0,w')=(-0.87,-0.48)$. The points $\at_{1,2,3}$ are the roots of the Jordan Ricci $R(\tilde{a})$. The
    ansatz $F_A(\tilde{a})$, with the solutions \textbf{PS I}, \textbf{PS II}, and \textbf{PS III}, coincides with 
    $F(\tilde{a})$ in the neighborhoods of $\at_1, \at_2,$ and $\at_3$ respectively.}
  \label{fig:8}
\end{figure}

\renewcommand{\arraystretch}{1.3}
\begin{table}
  \centering
  \begin{tabularx}{\textwidth}{|p{1.7cm}|X|p{2.2cm}|c|p{1.7cm}|}
    \hline
    Parameter\newline sets & Solutions for $\epsilon_{0,1,2}$ & Roots of\newline $R(\tilde{a})=0$ & Range of $\tilde{a}$, $\Delta\leq .01$ & Stability as\newline$\tilde{a} \to \tilde{a}_{1,2,3}$ \\
    \hline \hline
    Parameter \newline set I & $\epsilon_0 = 0.9136$,\newline $\epsilon_1 = 0.0015 {(\kappa^2 \rho_0)}^{-1}$,\newline  $\epsilon_2 = 0.0011{(\kappa^2 \rho_0)}^{-2}$ & $\at_1 = 0.7771 $ & $0.76607<\tilde{a}<0.83022$ & $F>0$,\newline $F'>0$\\
    \hline 
    Parameter \newline set II & $\epsilon_0 = 2.45336$ \newline $\epsilon_1 = - 4.33028 {(\kappa^2 \rho_0)}^{-1}$ \newline $\epsilon_2 = 28.65243 {(\kappa^2 \rho_0)}^{-2}$ & $\at_2 = 2.8525$ & $2.70185 < \tilde{a} < 3.09667 $ & $F>0$, \newline $F'<0$\\
    \hline
    Parameter \newline set III & $\epsilon_0 = 553.26752$ \newline $\epsilon_1 = 1.13965 {(\kappa^2 \rho_0)}^{-1}$ \newline  $\epsilon_2 = 1.06951  {(\kappa^2 \rho_0)}^{-2}$  & $\at_3 = 74.5742$ & $72.6205 < \tilde{a} < 82.2392 $ & $F>0$, \newline $F'>0$\\
    \hline
  \end{tabularx}
  \caption{Solutions for the ansatz $F_A$~\eqref{eq:42}, using the quintessence parameter
    values~\eqref{eq:43}. Different sets of solutions for $\epsilon_{0,1,2}$ are given in the second
    column. The third column shows the roots of $R(\tilde{a})=0$, $\tilde{a}_{1,2,3}$, in the neighbourhood of
    which the corresponding parameter set is applicable (also see Fig.~\ref{fig:8}). We define the
    specific deviations of the ansatz $F_A(\tilde{a})$ from the exact $F(\tilde{a})$ corresponding to a
    particular parameter set as
    $\Delta(\tilde{a},\ \text{PS}_{\text{I, II, III}})=\frac{|F_A(\tilde{a};\ \text{PS}_{\text{I, II, III}}) -
      F(\tilde{a})|}{|F(\tilde{a})|}$. In the fourth column, ranges of $\tilde{a}$ are shown for which $\Delta <
    0.01$. The fifth column shows whether $F, F'>0$ as $\tilde{a} \to \tilde{a}_{1,2,3}$.}
  \label{tab:1}
\end{table}
Each of the solutions for the ansatz $F_A$~\eqref{eq:42}, i.e., \textbf{Parameter set I} (\textbf{PS
  I}), \textbf{Parameter set II} (\textbf{PS II}), \textbf{Parameter set III} (\textbf{PS III}), is
applicable in the neighborhood of one of the points $\tilde{a}_{1,2,3}$, where $R(\tilde{a}) \to 0$
(see table~\ref{tab:1}). We compare the exact $F(\tilde{a})$ (from~\eqref{eq:45})
with $F_A(\tilde{a})$ in these neighborhoods in Fig.~\ref{fig:8}.

The best estimation for the ansatz, $F_A$, consistent with the current Einstein frame scale factor,
$\tilde{a} = 1$, is given by the \textbf{PS I} solutions, since $\tilde{a}_1$ is the nearest root of $R(\tilde{a})$ to
the current scale factor $\tilde{a} = 1$.  One can estimate the current value of Jordan Ricci using the
quintessence parameters~\eqref{eq:43},
\begin{align}
  \label{eq:46}
  R_0 = R(\tilde{a} = 1) = 1.3684 \kappa^2 \rho_0.
\end{align}
In order to see whether the ansatz~\eqref{eq:42}, with \textbf{PS I} solutions, is consistent at
the current universe ($\tilde{a} \to 1$), we derive the relative contributions of the terms in
$F_A(R)$. Using \textbf{PS I} from table~\ref{tab:1} and~\eqref{eq:46} we derive the three terms in
the ansatz $F_A(R_0)$,
\begin{subequations}
  \label{eq:47}
  \begin{align}
    \text{Term}_0 &= \epsilon_0 = 0.91357,\\
    \text{Term}_1 &= \epsilon_1 R_0 = 0.00207,\\
    \text{Term}_2 &= \epsilon_2 R_0^2 = 0.00205.
  \end{align}
\end{subequations}
As we can see, the zeroth order term ($\epsilon_0$) contributes in the order of $\sim 10^2$ times
than the first order term ($\epsilon_1R_0$), however, the contributions of the first order term and
the second order term $\epsilon_2 R_0^2$ are in the same order of magnitude. The specific deviation
of the ansatz $F_A(\tilde{a})$ from the exact $F(\tilde{a})$ at current scale factor is
\begin{align}
  \label{eq:48}
  \Delta(\tilde{a} = 1,\ \text{PS}_{\text{I}})=\frac{|F_A(\tilde{a} = 1;\ \text{PS}_{\text{I}}) - F(\tilde{a} =
    1)|}{|F(\tilde{a} = 1)|} = 0.08236.
\end{align}

In the following part we discuss the stability of these perturbative solutions.

\paragraph{Stability of perturbative solutions.}

For the perturbative solutions $F_A(R)$, using quintessence parameters~\eqref{eq:43} we numerically
find that $F_A>0$ in the limit $\tilde{a} \to \tilde{a}_{1,2,3}$, that is, for solutions
corresponding to \textbf{PS I, PS II, PS III} (see
table~\ref{tab:1}).  Using the \textbf{PS I} solutions for current Einstein frame
universe ($\tilde{a} \to 1$), we find that $F_A>0$ in the current epoch as well. We also find
that $F_A' = \epsilon_1 + 2 \epsilon_2 R$ is positive while $\tilde{a} \to \tilde{a}_{1,3}$,
however, $F_A'<0$ as $\tilde{a} \to\tilde{a}_2$. For the present day universe ($\tilde{a} = 1$), we
find $F_A'>0$ with the \textbf{PS I} solutions.

\subsubsection{Asymptotic solution for $F(R)$ in the late time}
\label{sec:asympt-solut-fr}
We first consider the late time behaviour of the scale factor in the Einstein frame.  According
to our choice of parameterization~\eqref{eq:14}, $w(\at)$ increases monotonically with the scale
factor $\at$. When $\at > \exp \left((1 + 3w_0)/(3w') \right)$, $w(\at)> -1/3$, and the quintessence
field can no longer drive the acceleration in the Einstein frame. From the Friedmann equation in
the Einstein frame,
\begin{align}
  \label{eq:49}
  \tilde{H} = \frac{1}{\tilde{a}} \diff{\tilde{a}}{\te} = \frac{\kappa^2}{3} \rho(\tilde{a}),
\end{align}
together with the solution of the continuity
equation, $\rho(\tilde{a}) = \rho_0 \tilde{a}^{-3(1+w_0)+ \frac{3}{2} w' \ln \tilde{a}}$, it is
evident that the $\dif \at/ \dif \te$ always remains positive. Therefore, even though the
acceleration in the Einstein frame stops at some finite time, the scale factor $\at$ keeps
increasing indefinitely. Solving the Friedmann equation we find the relation between the
coordinate time $\te$ and the scale factor to be
\begin{align}
  \label{eq:50}
  \te &=  \sqrt{\frac{3}{\kappa^2 \rho_0}} \frac{\sqrt{\pi}}{2 \sqrt{A_2}} \e^{- \frac{B^2_2}{4 A_2}} \erfi \left( \sqrt{A_2} \ln \tilde{a} - \frac{B_2}{2 \sqrt{A_2}} \right) + \text{constant},
\end{align}
where $A_2 = - 3w'/4 > 0$, $B_2 = - 3(1 + w_0)/2$ (recall that $w'<0$ is taken through out in order for
the field $\varphi$ to be real valued in the $\tilde{a} \to \infty$ limit).  It is evident
from~\eqref{eq:50} that $\tilde{a}$ monotonically increases with $\te$, hence the late time limit in
Einstein frame, defined as $\te \to \infty$, can also be expressed as $\tilde{a} \to \infty$.

From~\eqref{eq:45} one can see that $F(\tilde{a}) \to \infty$ in the late time limit of the Einstein frame
($\tilde{a} \to \infty$). Now let us consider the late time behaviour of Jordan frame Ricci $R(\tilde{a})$.
Starting with~\eqref{eq:34},
\begin{align}
  \label{eq:51}
  \frac{R(\fb)}{C} =  \exp \left(\fb + C_2{\mathcal{P}}^{\frac{4}{3}}\right)\left[ 4 - 2{\mathcal{P}}^{\frac{2}{3}} + \frac{2}{3}B {\mathcal{P}}^{-\frac{1}{3}}-\frac{8}{3} B C_2 {\mathcal{P}}^{\frac{1}{3}}+ \frac{4}{3} B C_2 \mathcal{P}\right],
\end{align}
we see, $\fb = \ln F \to \infty$ in the late time limit, thus the exponential factor in the last
equation is dominant in this limit. Moreover, the second term in the argument of the exponential is
dominant, which is,
\begin{align}
  C_2 \mathcal{P}^{\frac{4}{3}} = C_2 (A- B \fb)^{\frac{4}{3}} \approx C_2(-B \fb)^{\frac{4}{3}} \to - \infty\ \text{as}\ \fb \to \infty.
\end{align}
Note that since $w'<0$, from~\eqref{eq:35}, $B,C_2<0$. Using this and taking the late time limit
of~\eqref{eq:51}, one can conclude that Jordan frame Ricci $R \to 0^+$ in the limit $\tilde{a} \to \infty$
(see Fig.~\ref{fig:3}). Hence, even if in the limit $\tilde{a} \to \infty$ we see $R \to 0^+$, we also
find $F \to \infty$, which makes the ansatz $F_A$~\eqref{eq:42} insufficient in the late time limit
of Einstein frame.

It is possible to derive an asymptotic functional form of $F(R)$ consistent with the late time limit
of Einstein frame. Noting that $F, \fb \to \infty$ in this limit, we consider only the most dominant
terms in each of the factors of $R(\fb)$ in~\eqref{eq:51} to obtain
\begin{align}
  \label{eq:52}
  \lim_{\fb \to \infty} \frac{R(\fb)}{C} &\approx N \fb \exp \left( - M \fb^{\frac{4}{3}} \right),
\end{align}
where, 
\begin{subequations}
  \label{eq:53}
  \begin{align}
    B, C_2 &<0,\\
    N &= - \frac{4}{3} B^2 C_2 > 0,\\
    M &= -C_2 (-B)^{\frac{4}{3}} > 0.
  \end{align}
\end{subequations}
Taking $\ln$ of~\eqref{eq:52} we get
\begin{align}
  \label{eq:54}
  \ln \left( \frac{R}{C} \right) &= \ln (N \fb) - M \fb^{\frac{4}{3}},
\end{align}
neglecting the $\ln$ term in RHS,
\begin{align}
  \label{eq:55}
  \fb_{LT} &= {\left[ - \frac{1}{M} \ln \left( \frac{R}{C} \right)  \right]}^{\frac{3}{4}},\\
  \label{eq:56}
  \text{and }F_{LT}(R) &= \exp \left( {\left[ - \frac{1}{M} \ln \left( \frac{R}{C} \right) \right]}^{\frac{3}{4}} \right).
\end{align}
The subscript $LT$ represents the late time of Einstein frame. Hence, as long as $ R/C < 1$ ($C>0$,
as defined in~\eqref{eq:35}), $F_{LT}$ remains real, which is acceptable since this is valid
in large $F$ and small $R$ limit.  We can plug in $\fb$ from~\eqref{eq:55} in~\eqref{eq:33} to get
an asymptotic expression for $f(R)$.
\begin{subequations}
  \label{eq:57}
  \begin{align}
    \label{eq:58}
    f(\fb) &= C \exp\left( 2 \fb + C_2 \mathcal{P}^{\frac{4}{3}} \right) \left[ 2 - \mathcal{P}^{\frac{2}{3}} + \frac{2}{3} B \mathcal{P}^{-\frac{1}{3}} - \frac{8}{3} B C_2 \mathcal{P}^{\frac{1}{3}} + \frac{4}{3} B C_2 \mathcal{P}  \right]\\
    \label{eq:59}
           &\approx C \exp \left( C_2(-B)^{\frac{4}{3}} \fb^{\frac{4}{3}} \right) \left[ -\frac{4}{3}B^2 C_2 \fb \right]\\
    \label{eq:60}
    f_{LT}(R) &= N R \left[ - \frac{1}{M} \ln \left( \frac{R}{C} \right) \right]^{\frac{3}{4}}
  \end{align}
\end{subequations}
where in the second line, we have considered only the highest order terms in
$\fb$ in the argument of the exponential and in the square bracket.

It is evident from~\eqref{eq:56} that for the asymptotic solution $F_{LT}>0$, as long as
$0 < R/C <1$, this satisfies the `positive gravitational coupling' condition $F>0$. However,
$F_{LT}'<0$, where
\begin{align}
  \label{eq:61}
  F_{LT}' = - \frac{3}{4 M R} \frac{1}{\left[ - \frac{1}{M} \ln \left( \frac{R}{C} \right)
    \right]^{\frac{1}{4}}} \exp \left( \left[- \frac{1}{M} \ln \left( \frac{R}{C} \right)
    \right]^{\frac{3}{4}}\right).
\end{align}
The late time solution $F_{LT}$ seemingly violates the `matter instability condition' as discussed
previously. However, the `Dolgov-Kawasaki/matter instability' condition is not inherently suitable
in the late time limit of the current $f(R)$ model. The derivation of the stability condition
assumes that the $f(R)$ theory is a small deviation of Einstein gravity~\eqref{eq:36}
(see~\cite{Faraoni_2006,sotiriou2010} for the derivation), this implies
$F(R) = f'(R) = 1 + \varepsilon\psi'(R)$. However, for the current model in the late time limit, we
see that $F_{LT}(\tilde{a} \to \infty) \to \infty$, it cannot be considered as a perturbation on
Einstein gravity.

\section{Relation between Jordan and Einstein frame scale factors}
\label{sec:comparison-late-time}

\begin{figure}
  \centering
  \includegraphics[width=.7\textwidth]{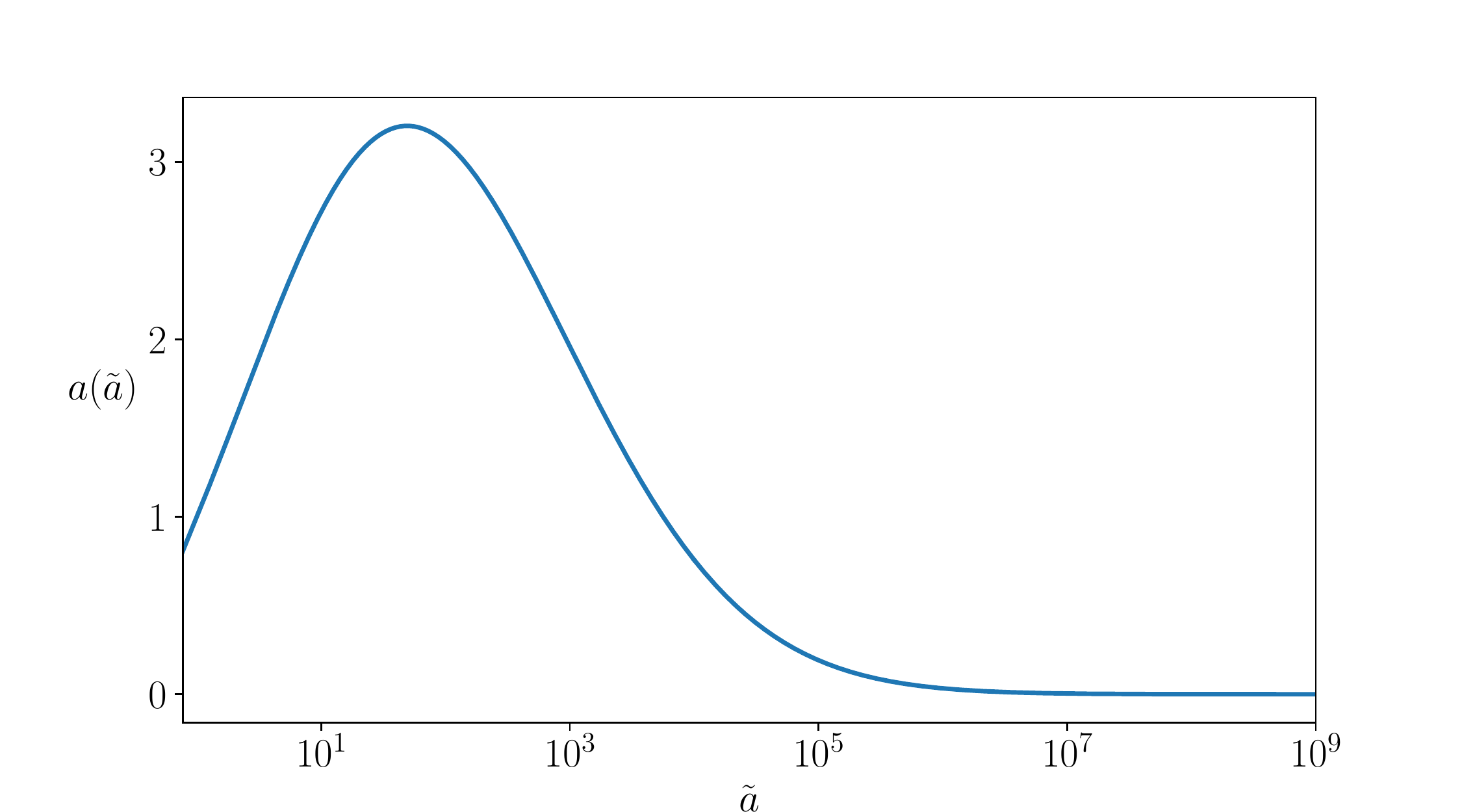} 
  \caption{Late time collapse of the Jordan frame universe. For a quintessence model with logarithmic
    equation of state parameter, $w(\tilde{a}) = w_0 - w' \ln \tilde{a},\ w'<0$. The Jordan frame
    scale factor plotted as a function of the Einstein frame scale factor, $a(\tilde{a})$, with
    quintessence parameters $(w_0,w') = (-0.87, -0.48)$. $a(\tilde{a})$ has a maximum at
    $\tilde{a} = \exp ((w_0 - 1)/w') = 49.197$. In the late time limit $a \to 0$ as
    $\tilde{a} \to \infty$.}
  \label{fig:9}
\end{figure}

In this section we explore the late time relation between the Jordan and Einstein frame scale
factors. For the current model, one can obtain the Einstein frame Ricci scalar as
\begin{align}
  \label{eq:62}
  \tilde{R} (\tilde{x}) = \kappa^2 \rho_0 ( 1 - 3w_0 + 3 w' \tilde{x}) \exp{\left[ -3(1+w_0) \tilde{x} + \frac{3}{2} w' \tilde{x}^2\right] },
\end{align}
where $\tilde{x} = \ln \tilde{a}$.  We can see that the Einstein frame universe becomes
Ricci flat in the late time limit, i.e., $\tilde{R} \to 0$ as $\tilde{x} = \ln \tilde{a} \to \infty$
(recall that we consider $w'<0$ as discussed before). We have already seen that the Jordan frame
Ricci scalar $R$ vanishes in this limit too. However, as the Einstein frame scale factor
$\tilde{a} \to \infty$ in the late time limit (see~\eqref{eq:50}), the Jordan frame scale factor
$a \to 0$, resulting in a collapsing Jordan frame universe.  To see this, consider the relation
between the line elements in Einstein ($\dif \tilde{s}^2$) and Jordan ($\dif s^2$)
frames~\cite{defelice2010},
\begin{subequations}
  \label{eq:63}
  \begin{align}
    \label{eq:64}
    \dif \tilde{s}^2 &= - \dif \te^2 + \tilde{a}^2 \dif \mathbf{\tilde{x}}^2 \\
    \label{eq:65}
                     &= \Omega^2 \dif s^2\\
    \label{eq:66}
                     &= F (-\dif t^2 + a^2 \dif \mathbf{x}^2),
  \end{align}
\end{subequations}
which implies
\begin{align}
  \label{eq:67}
  \dif \te &=  \sqrt{F} \dif t\\
  \label{eq:68}
  a &= \frac{\tilde{a}}{\sqrt{F}}.
\end{align}
Using this the Jordan frame scale factor can be written in terms of the Einstein frame scale factor as
\begin{align}
  \label{eq:69}
  a( \tilde{a} ) = \exp(C_5) \tilde{a} \exp \left( \frac{\sqrt{2}}{3w'} (1 + w_0 - w' \ln \tilde{a})^{\frac{3}{2}} \right),
\end{align}
where $C_5 = -\frac{\sqrt{2}}{3w'} (1 + w_0)^{\frac{3}{2}}$, equation~\eqref{eq:45} is used to derive
the last equation. $a(\tilde{a})$ reaches the maximum value at $\tilde{a} = \exp ((w_0 - 1)/w')$,
after which it decreases monotonically. For the quintessence parameters in~\eqref{eq:43}, $a(\at)$
becomes maximum as $\at \to 49.197$. In the late time limit of the Einstein frame
($\tilde{a} \to \infty$), the Jordan frame scale factor $a \to 0$ for $w'<0$ (see
Fig.~\ref{fig:9}).

This result leads to two interesting observations. First, the Jordan frame scale factor has a
maximum value at the critical point $w(\at) = 1$, this means the Jordan frame universe has a bound
on its spatial size, which is determined by the quintessence parameters ($w_0,w'$). This is in
contrast with the Einstein frame universe, as we have seen in~\eqref{eq:50}, the Einstein frame
scale factor grows indefinitely with the coordinate time. Second, the collapsing universe in the
Jordan frame provides an equivalent description of the expanding universe in the Einstein
frame. This map between collapsing and expanding frames can be a useful tool. For example,
introducing metric perturbations in the backgrounds of the FRW spacetime, one may establish a
relation between the perturbation in the expanding frame with the perturbation in the collapsing
frame. Perturbation in a collapsing gravitational system can be analyzed within the $f(R)$
setting~\cite{Cembranos12}. Further studies can reveal how the quantum effects in
the collapsing Jordan frame correspond to the perturbations in the ever-expanding Einstein frame at
late times (see, for example,~\cite{narlikar78,kuroda84,casadio00,koch13,martin20} and references
therein). These issues will be pursued elsewhere.

\section{Expansion-collapse duality in the presence of non-relativistic matter}
\label{sec:expans-coll-dual}
While discussing the expansion-collapse duality of the two conformally connected frames, we have so
far considered the quintessence field to be the only component in the Einstein frame universe. In
this section we consider a more realistic scenario where the Einstein frame universe consists of the
quintessence field as well as matter. Although dark energy is the dominant component of the current
epoch, even the subdominant presence of dust (non-relativistic pressureless fluid) staggers the
Jordan frame collapse.

Since we are considering the Einstein frame to be the physical frame; observations are made in this
frame. Therefore, leaning to the standard considerations, in the Einstein frame matter is minimally
coupled to gravity, there is no explicit coupling between matter and the quintessence field
$\varphi$. Since the Einstein and Jordan frames are conformally related, the couplings transform
non-trivially between the frames-- that matter will be non-minimally coupled to gravity in the
Jordan frame. Therefore, for a more realistic description of the Einstein frame, the total Jordan
frame action will have a form
\begin{align}
  \label{eq:73}
  \mathcal{S}_J &= \frac{1}{2\kappa^2} \int d^4 x \sqrt{-g}f(R) + \int d^4 x \sqrt{- g} F^2(R) \mathcal{L}_{\text{M}}(\psi_{\text{M}}; F(R) g_{\mu \nu}).
\end{align}
Since $F(R)=\partial f/\partial R$, the matter-gravity coupling is through some function of Ricci
scalar and belongs to a broad class of curvature couplings \cite{harko14}. This form of the
action in the Jordan frame is chosen such that, after conformal transformation
$\tilde{g} _{\mu \nu} = F g _{\mu \nu}$, matter decouples from the curvature $\tilde {R}$ in the
Einstein frame,
\begin{align}
  \label{eq:74}
  \mathcal{S}_E = \int d^4x \sqrt{-\tilde{g}} \frac{1}{2\kappa^2} \tilde{R}-\int d^4x \sqrt{-\tilde{g}} \left[ \frac{1}{2} \tilde{g}^{\mu \nu} \partial_\mu \varphi \partial_\nu \varphi + V(\varphi)\right] + \int d^4 x \sqrt{- \tilde{g}}  \mathcal{L}_{\text{M}}(\psi_{\text{M}};  \tilde{g} _{\mu \nu}).
\end{align}
Due to introduction of matter, solutions for the Jordan frame action obtained previously for the
case of pure gravity in Jordan frame will not be consistent with action~\eqref{eq:73}. Thus the
equation of motion, condition for acceleration and the relation between the two frames' scale factor
will be modified, {\it however we will see that the explicit form of the solution of this action
  does not affect the existence of the expansion-collapse duality. Thus, the existence of such
  duality between the frames can be considered to be a robust feature.}

From the Einstein frame action~\eqref{eq:74}, one can obtain the usual Friedmann equation
\begin{align}
  \label{eq:75}
  \tilde{H}^2 = \frac{\kappa^2}{3} \left( \rho_{m} + \rho_{\varphi} \right),
\end{align}
where $\tilde{H}$ is the Hubble parameter in the Einstein frame, $\rho_m$ and $\rho_{\varphi}$ are the
energy densities corresponding to dust and the quintessence field respectively. Together with
observationally consistent values of ($\Omega_{m0}, \Omega_{\varphi 0}, w_\varphi$),
Eq.~\eqref{eq:75} correctly reproduces the standard expansion history of the universe without
spatial curvature in the Einstein frame. The case with spatial curvature is discussed later.

\begin{figure}
  \centering
  \includegraphics[width=.7\textwidth]{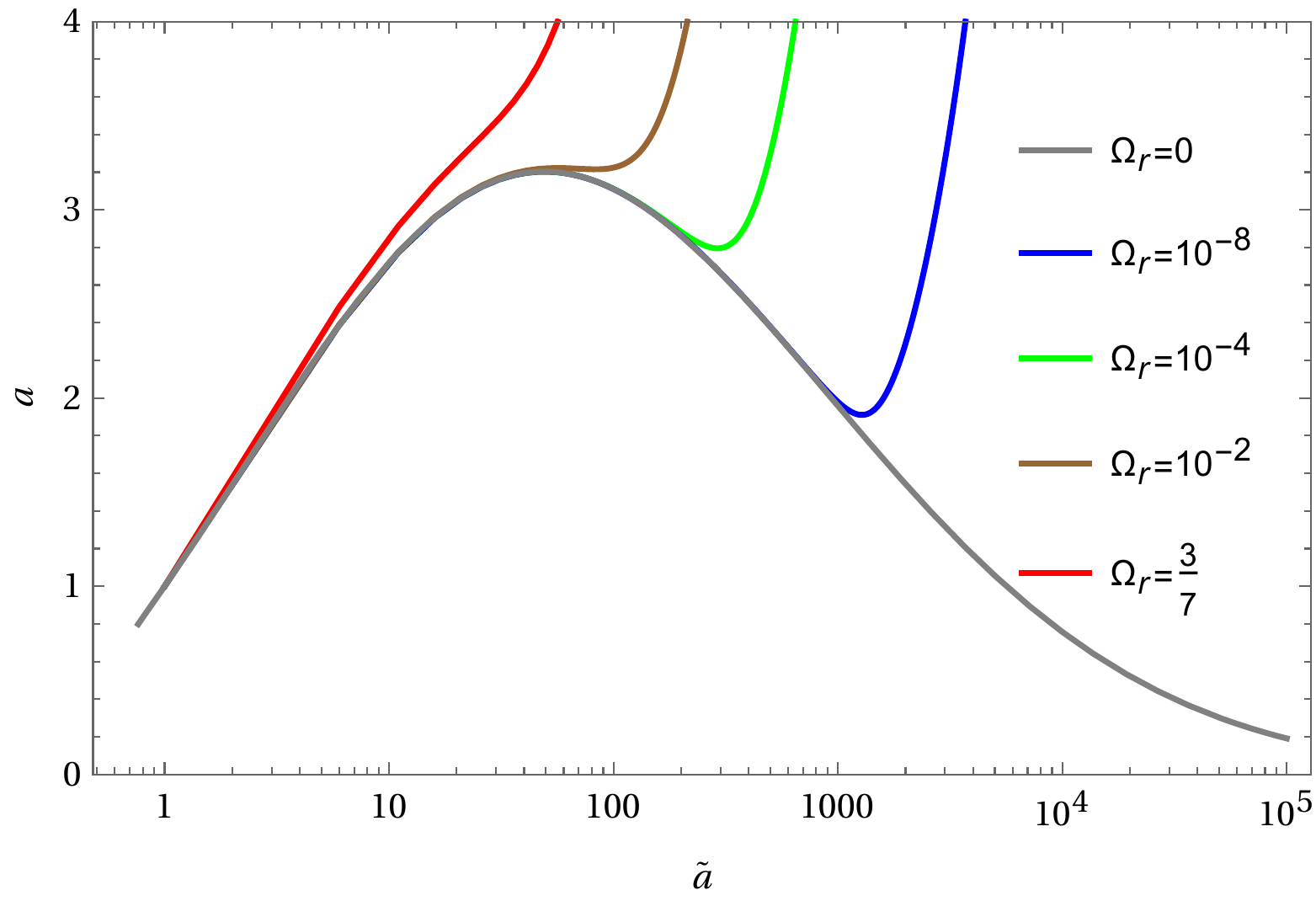} 
  \caption{Evolution of Jordan frame scale factor $a$ with respect to Einstein frame scale factor
    $\at$ for different values of $\Omega_r$, where $\Omega_r = \Omega_{m0}/\Omega_{\varphi 0}$ is
    the ratio of the density parameters of dust and the quintessence field at the current epoch
    $\at = 1$ and $(w_0, w') = (-0.87, -0.48)$. Any non zero value of $\Omega_{m 0}$ prevents the
    Jordan frame universe to collapse indefinitely. For $\Omega_r$ consistent with observations,
    $\Omega_r = \Omega_{m0}/\Omega_{\varphi 0} \sim 0.3/0.7$, the Jordan frame expands monotonically
    with respect to the Einstein frame.}
  \label{fig:10}
\end{figure}

Using~\eqref{eq:75}
in~\eqref{eq:6} for the Einstein frame we obtain
\begin{align}
  \label{eq:76}
  \diff{\Phi}{\at} = \frac{\sqrt{3}}{\kappa} \frac{1}{\at} \sqrt{\frac{1 + w_{\varphi}(\at)}{1 + \rho_m/\rho_{\varphi}}},
\end{align}
where $w_\varphi$ is the equation of state parameter of the quintessence field. This, together
with~\eqref{eq:68}, leads to a relation between the Jordan and the Einstein frame scale factors
($a,\ \at$),
\begin{align}
  \label{eq:77}
  a(\at) = \at \exp \left( - \frac{1}{\sqrt{2}} \int_{1}^{\at} \frac{d \at '}{\at '} \sqrt{\frac{1 + w_{\varphi}}{1 + \rho_m/\rho_{\varphi}}} \right).
\end{align}
For the logarithmic $w_\varphi$~\eqref{eq:14}, the above equation becomes
\begin{align}
  \label{eq:78}
  a(\at) = \at \exp \left( - \frac{1}{\sqrt{2}} \int_{1}^{\at} \frac{d \at '}{\at '} \sqrt{\frac{1 + w_0 - w' \ln \at'}{1 + \Omega_r \exp \left( 3 w_0 \ln \at' - \frac{3}{2} w' (\ln \at')^2 \right)}} \right),
\end{align}
where we have
used
$\rho_m = \rho_{m0} \at^{-3}$,
$\rho_\varphi = \rho_{\varphi 0} \at^{-3(1 + w_0) + \frac{3}{2} w' \ln \at}$,
$\Omega_r = \Omega_{m0}/\Omega_{\varphi 0} = \rho_{mo}/\rho_{\varphi 0}$,
and $\Omega_{m0, \varphi 0}$ are the density parameters corresponding to dust and the quintessence
field at $\at = 1$. For $\Omega_r=0$, that is, if the energy contribution of dust is completely
neglected, the equation~\eqref{eq:78} can be solved analytically, resulting in~\eqref{eq:75} as is
expected. However, for nonzero $\Omega_r$, equation~\eqref{eq:78} can be solved only
numerically. Fig.~\ref{fig:10} shows the evolution of the Jordan frame scale factor with respect to
the Einstein frame scale factor for different values of $\Omega_r$. We see that even if the energy
contribution due to dust with respect to that of the quintessence field is taken to be as small
as $\Omega_r = 10^{-8}$ at the current epoch, the Jordan frame collapse sustains only for a finite
duration of time. Eventually in the late times of the Einstein frame, Jordan frame scale factor turns
around and starts to increase monotonically. For the value of the dust-dark energy ratio consistent
with observation, $\Omega_r \sim 0.3/0.7$, we see that the Jordan frame universe never collapses,
the Jordan frame scale factor evolves monotonically with respect to the Einstein frame scale factor.

In fact, it is generally possible to write the condition for the Jordan frame collapse in a
relatively simple inequality. Starting with~\eqref{eq:68} we can obtain the derivative of Jordan
frame scale factor with respect to that of the Einstein frame,
\begin{align}
  \label{eq:79}
  \diff{a}{\at} = \diff{ }{\at} \left( \frac{\at}{\sqrt{F}} \right) = F^{-1/2} \left( 1 - \kappa \sqrt{1 + w_\varphi(\at)} \sqrt{\frac{\rho_\varphi}{6 \tilde{H}^2}}\right),
\end{align}
where we have used the Einstein-Jordan frame correspondence, $2 \kappa \varphi = \sqrt{6} \ln F$, along with the result
\begin{align}
  \diff{\varphi}{\at} = \frac{1}{\at} \sqrt{1 + w_\varphi(\at)} \sqrt{\frac{\rho_\varphi}{\tilde{H}^2}}.
\end{align}
It is evident from~\eqref{eq:79} that the Jordan frame universe collapses, that is,
$\dif a/\dif \at < 0$ if and only if the following inequality is satisfied,
\begin{subequations}
  \label{eq:80}
  \begin{align}
    \label{eq:81}
    w_\varphi(\at) &> \mathcal{C}(\at)\\
    \label{eq:82}
    \mathcal{C}(\at) &= \frac{6 \tilde{H}^2}{\kappa^2 \rho_\varphi(\at)} - 1.
  \end{align}
\end{subequations}
The above condition leads to a range (or possibly multiple ranges) of $\at$ during which the Jordan
frame collapses. This also shows that in order for the Jordan frame to have a `turn around', that is
where $\dif a/\dif \at = 0$, the equation $w_\varphi(\at) = \mathcal{C}(\at)$ must have a solution.
This is applicable for arbitrary quintessence models; it also takes into account the presence of any
other component in the universe, like dust, radiation or spatial curvature. However, it is to be
noted that the condition is only imposed on $w_\varphi$, not on the effective equation of state
parameter of the era and the existence of all the other components is incorporated in the Hubble
parameter $\tilde{H}$, through the Friedmann equation. We also point out that this inequality has a
significantly simple form. Starting with any given $w_\varphi(\at)$, one can check whether the
condition is satisfied just by knowing $\rho_\varphi(\at)$. The functional form of $f(R)$, the
solution of the quintessence field $\varphi(\at)$ or even the form of the potential $V(\varphi)$ are
not required for that matter. Further, we see from Eq. (\ref{eq:79}) the condition $ da/d\tilde{a}<0$ is insensitive to exact from of
$F(R)$ which also plays the role of non-minimal coupling in the Jordan frame, but is solely decided by the ratio $6 \tilde{H}^2/\kappa^2 \rho_\varphi(\at)$, i.e. on the phenomenological form of the quintessence field arrived at in the Einstein frame.  Thus, the onset of collapse in the Jordan frame remains persistent even in the presence of the proposed non-minimal coupling.

For the quintessence model under discussion, the inequality~\eqref{eq:81} takes the form
\begin{subequations}
  \label{eq:83}
  \begin{align}
    \label{eq:84}
    w_\varphi(\at) &> \mathcal{C}(\at) \text{, where}\\
    \label{eq:85}
    w_\varphi(\at) &= w_0 - w' \ln \at,\\
    \label{eq:86}
    \mathcal{C}(\at) &= 1 + 2 \frac{\Omega_{m0}}{\Omega_{\varphi 0}} \exp \left( 3 w_0 \ln \at - \frac{3}{2} w' (\ln \at)^2\right).
  \end{align}
\end{subequations}
in the presence of matter. For the values of the parameters $(w_0,w',\Omega_{m0})$ consistent with
observations, one can argue that the above inequality is never satisfied (see
Appendix~\ref{sec:appendix2} for the argument), thus the quintessence model with logarithmic
equation of state parameter does not lead to a collapsing Jordan frame in the presence of dust.

\begin{figure}
  \centering
  \includegraphics[width=.7\textwidth]{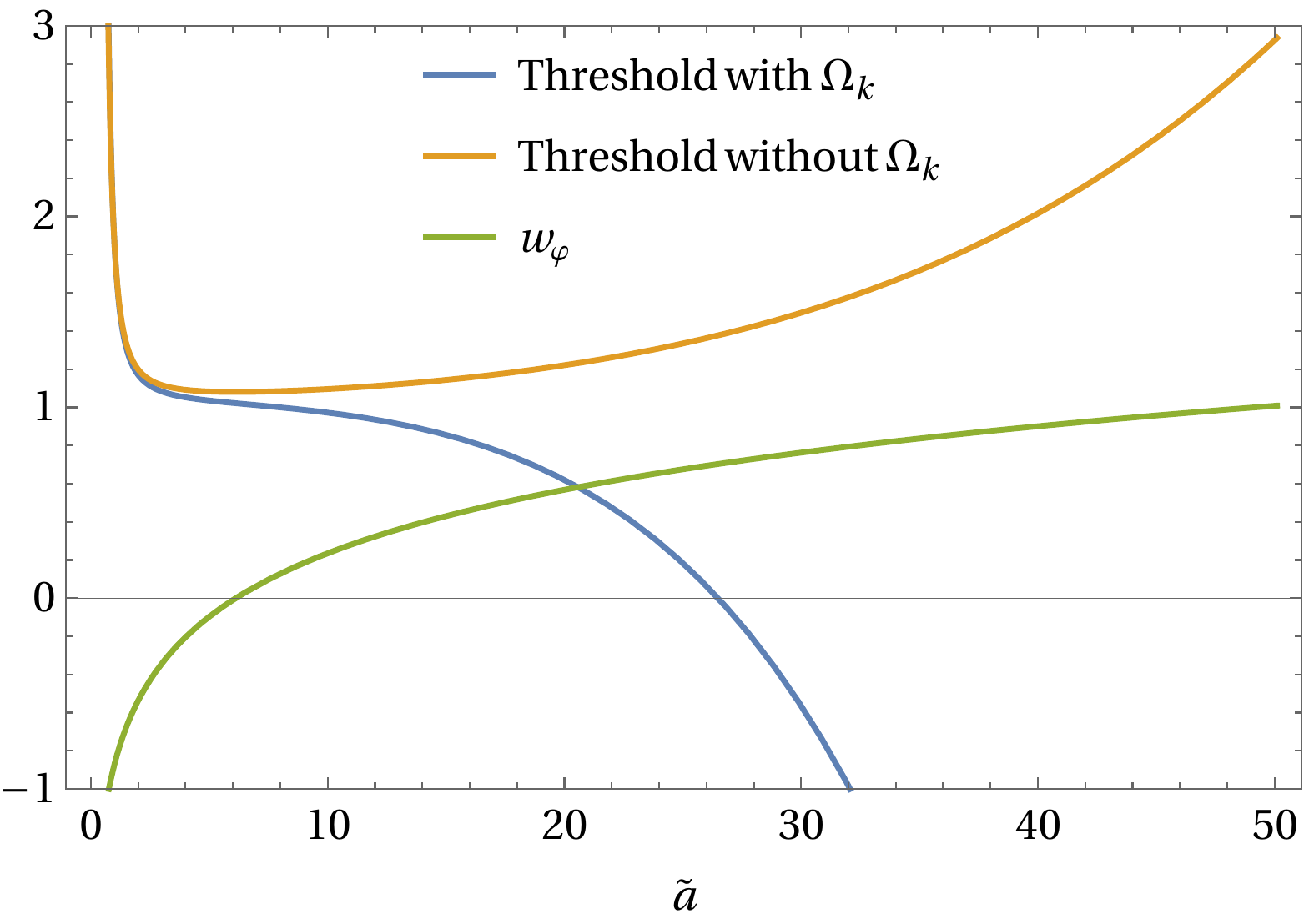} 
  \caption{Possibility of Jordan frame collapse in the presence of positive spatial curvature and
    dust, with
    $(w_0, w') = (-0.87,-0.48),\ \Omega_{m0} = 0.315,\ \Omega_{\varphi 0} = .728,\ \Omega_{K 0} =
    -.043$. `Threshold' refers to the function $\mathcal{C}$ in~\eqref{eq:88}. The Jordan frame
    collapse while $w_\varphi (\at)>\mathcal{C}(\at)$ is satisfied. The plot shows that in contrast
    with the case when $\Omega_k=0$, the negative contribution from $\Omega_K$ brings down the
    threshold $\mathcal{C}$ eventually. This allows for $w_\varphi$ to cross the threshold at
    $\at \sim 20.52$, after which the Jordan frame starts to collapse. }
  \label{fig:11}
\end{figure}

We conclude this section by discussing the effect of a positive spatial curvature on the
expansion-collapse duality. As it is argued in~\cite{Valentino_2019}, the latest \textit{Planck
  data}~\cite{planck_2018} appears to be consistent with a universe possessing a small positive
spatial curvature. We find that the introduction of a positive spatial curvature in the analysis
above recovers the expansion collapse duality in the presence of dust. Taking into account both
spatial curvature and dust,  the Friedmann equation in the Einstein frame becomes
\begin{align}
  \label{eq:70}
  \tilde{H}^2 = \frac{\kappa^2}{3} \left( \rho_\varphi + \rho_m \right) - \frac{K}{\at^2},
\end{align}
where $K$ is the spatial curvature parameter appearing in the FRW metric. In this case the
condition of expansion-collapse duality~\eqref{eq:81} becomes
\begin{subequations}
  \label{eq:87}
  \begin{align}
    \label{eq:88}
    w_\varphi(\at) &> \mathcal{C} (\at) \text{, where}\\
    \label{eq:89}
    w_\varphi(\at) &= w_0 - w' \ln \at, \\
    \label{eq:90}
    \mathcal{C}(\at) &= 1 + 2 \frac{\Omega_{m0}}{\Omega_{\varphi 0}}  \exp \left( 3 w_0 \ln \at - \frac{3}{2} w' (\ln \at)^2\right) \nonumber \\
                   &+ 2 \frac{\Omega_{K 0}}{\Omega_{\varphi 0}} \exp \left( (1 +3 w_0) \ln \at - \frac{3}{2} w' (\ln \at)^2\right),
  \end{align}
\end{subequations}
where $\Omega_K = - K/(\at \tilde{H})^2 $, $\Omega_{K 0} = -K/\tilde{H}_0^2$ is the current value of the
density parameter associated with the spatial curvature. As we can see from the condition,
introduction of a positive spatial curvature, that is, a negative $\Omega_K$, brings down the
threshold that $w_\varphi$ is required to cross in order for the Jordan frame collapse to begin. For
example, with the value $\Omega_{K0} = -.0438$~\cite{Valentino_2019}, Fig.~\ref{fig:11} shows region
of $\at$ where the condition for Jordan frame collapse is satisfied in the presence of matter.

To summarize, we show that the quintessence model with logarithmic equation of state parameter leads
to an expansion-collapse duality between Einstein and Jordan frames, that is, while the Einstein
frame expands indefinitely, the Jordan frame collapses in the late time. We generalize the
requirement of this expansion-collapse duality for arbitrary quintessence models, taking into
account the presence of other components in the Einstein frame as well. This requirement essentially
imposes a threshold on the equation of state parameter of the quintessence field, the threshold
itself is determined by the form of energy densities associated with all the components in the
Einstein frame universe. Using this we find that the Jordan frame collapse gradually becomes short
lived if dust is included in the Einstein frame universe. However, the Jordan frame collapse, at
least for a finite duration, can be recovered by introducing a small positive spatial curvature.

\section{Conclusion and discussion}
\label{sec:concl-disc}

Quintessence fields are viable alternatives to the $\Lambda$CDM models, which can provide an
explanation for the recent accelerated expansion of the universe. It is known that
an $f(R)$ theory of gravity in the Jordan frame acts as a dual to a quintessence field in Einstein
gravity, in a conformally connected spacetime, known as the Einstein frame. In this work, we
reconstruct $f(R)$ functions in Jordan frame which reproduce quintessence models with
time-independent and time-dependent equation of state parameter ($w$) in the Einstein frame.

As an example of time-dependent equation of state parameter, we choose the logarithmic
parameterization and reconstruct $f(R)$ function corresponding to the quintessence model. This
solution is obtained in two parts. A perturbative solution of $F(R)$ is obtained which is valid in
the small curvature limit of the Jordan frame. We show that this perturbative solution may be
applicable in the near future of the Einstein frame universe. However, we find that this ansatz
becomes ill-defined in the late time limit of Einstein frame. We obtain an asymptotic solution
for $f(R)$, which is valid in large field and small curvature limit of the Jordan frame. This
asymptotic solution is applicable in the late time limit of the Einstein frame. We also show that
the Jordan frame scale factor has a finite maximum value determined by the quintessence model
parameters, after which it keeps on decreasing. In the late time limit of the Einstein frame, the
Jordan frame universe collapses while the expansion in the Einstein frame universe continues.

We generalize this result and obtain the condition for expansion-collapse duality in terms of a
simple inequality. This condition can predict the possibility of Jordan frame collapse for
quintessence models with arbitrary equation of state parameters, even in the presence of other
components in the universe such as dust, radiation or spatial curvature. Given an equation of state
parameter, such prediction can be made only from the knowledge of energy density of the quintessence
model. Using this condition we show that the example quintessence model does not necessarily lead
to the Jordan frame collapse in the presence of dust, however the expansion-collapse duality may be
recovered by introducing a positive spatial curvature component in the universe.

The general condition for expansion-collapse duality can be readily applied to other physically
viable quintessence models. This opens a possibility of further studies of the collapse of the
Jordan frame. The mapping between expanding and collapsing geometries may have implications on
growth of perturbations which is a subject of further exploration.

\section*{Acknowledgement}

Research of K.L. is partially supported by the DST, Government of India through the DST INSPIRE
Faculty fellowship (04/2016/000571). The authors acknowledge the useful discussions with Jasjeet
S. Bagla.


\appendix
\section{Analytical results for the perturbative solution $F_A$}
\label{sec:appendix1}
Here we present the solutions of the ansatz~\eqref{eq:42}, i.e., we write the constants
$\epsilon_0, \epsilon_1, \epsilon_2$ in terms of the quintessence parameters $w_0,w'$. We use the
ansatz~\eqref{eq:42} in the RHS of~\eqref{eq:34} and ignore terms of the order $R^2$ and higher to
obtain
\begin{align}
  \label{eq:91}
  \begin{split}
    R =\ &C \alpha_0 \beta_0 \epsilon_0 + C \left[ \alpha_1 \beta_0 \epsilon_0 + \alpha_0 \beta_0 \epsilon_1 + \alpha_0 \beta_1 \epsilon_0 \right] R \\
    &+\ C \left[ \alpha_2 \beta_0 \epsilon_0 + \alpha_1 \beta_0 \epsilon_1 + \alpha_0 \beta_0 \epsilon_2 + \alpha_1 \beta_1 \epsilon_0 + \alpha_0 \beta_1 \epsilon_1 + \alpha_0 \beta_2 \epsilon_0 \right]R^2,
  \end{split}
\end{align}
where we have defined the following terms
\begin{subequations}
  \label{eq:92}
  \begin{align}
    \label{eq:93}
    k_0 &= A - B \ln \epsilon_0 \\
    \label{eq:94}
    k_1 &= -B \frac{\epsilon_1}{\epsilon_0} \\
    \label{eq:95}
    k_2 &= -B \left(\frac{\epsilon_2}{\epsilon_0}-\frac{\epsilon_1^2}{2 \epsilon_0^2} \right),
  \end{align}
\end{subequations} 
\begin{subequations}
  \label{eq:96}
  \begin{align}
    \label{eq:97}
    \alpha_0 &= e^{C_2 k_0^{\frac{4}{3}}}\\
    \label{eq:98}
    \alpha_1 &=  \frac{4}{3}C_2k_0^{\frac{1}{3}}k_1 e^{C_2
               k_0^{\frac{4}{3}}} \\
    \label{eq:99}
    \alpha_2 &= \left[ \frac{4}{3}C_2 k_0^{\frac{1}{3}} k_2 + \frac{2}{9}C_2 k_1^2 k_0^{-\frac{2}{3}} + \frac{8}{9} C_2^2 k_0^{\frac{2}{3}} k_1^2 \right] e^{C_2 k_0^{\frac{4}{3}}},
  \end{align}
\end{subequations}
and
\begin{subequations}
  \label{eq:100}
  \begin{align}
    \label{eq:101}
    \beta{}_0 = &+ 4 - \frac{8}{3}B C_2 k_0^{\frac{1}{3}} + \frac{2}{3}Bk_0^{-\frac{1}{3}} -2
                  k_0^{\frac{2}{3}} + \frac{4}{3} B
                  C_2 k_0\\
    \label{eq:102}
    \beta{}_1 = &-\frac{8}{9} B C_2 k_1 k_0^{-\frac{2}{3}} -\frac{2}{9} B k_1 k_0^{-\frac{4}{3}}
                  -\frac{4}{3} k_1 k_0^{-\frac{1}{3}} +
                  \frac{4}{3} B C_2 k_1 \\
    \label{eq:103}
    \beta_2 = &- \frac{8}{9}B C_2 k_2 k_0^{-\frac{2}{3}} + \frac{8}{27}B C_2 k_1^2 k_0^{-\frac{5}{3}} -\frac{2}{9}B k_2 k_0^{-\frac{4}{3}} +\frac{4}{27}B k_1^2 k_0^{-\frac{7}{3}} \nonumber\\
                &- \frac{4}{3}k_2 k_0^{-\frac{1}{3}} + \frac{2}{9}k_1^2k_0^{-\frac{4}{3}} +\frac{4}{3}B C_2 k_2.
  \end{align}
\end{subequations}
The terms $A, B, C_1, C_2, C$ were defined in~\eqref{eq:35}. Now we can compare the coefficients of
different powers of $R$ in~\eqref{eq:91} to determine the $\epsilon_{0,1,2}(w0,w')$. Comparing the
coefficients of $R^0$ we find
\begin{align}
  \label{eq:104}
  f_0(\epsilon_0) = \alpha_0 \beta_0 \epsilon_0 = 0.
\end{align}
Form~\eqref{eq:96} we seen $\alpha_0 \neq 0$,~\eqref{eq:104} becomes
\begin{subequations}
  \label{eq:105}
  \begin{align}
    \label{eq:106}
    \beta_0 &= 0\\
    \label{eq:107}
    4 - \frac{8}{3}B C_2 k_0^{\frac{1}{3}} + \frac{2}{3}Bk_0^{-\frac{1}{3}} -2 k_0^{\frac{2}{3}} + \frac{4}{3} B C_2 k_0 &= 0\\
    \label{eq:108}
    4 k_0^{\frac{1}{3}} - \frac{8}{3}B C_2 k_0^{\frac{2}{3}} + \frac{2}{3}B -2 k_0 + \frac{4}{3} B C_2 k_0^{\frac{4}{3}} &= 0\\
    \label{eq:109}
    \frac{3}{BC_2} k_0^{\frac{1}{3}} - 2 k_0^{\frac{2}{3}} + \frac{1}{2C_2} - \frac{3}{2BC_2} k_0 +  k_0^{\frac{4}{3}} &= 0\\
    \text{\{putting $k_0^{\frac{1}{3}} = x$, $(BC_2)^{-1} = m$, $(2C_2)^{-1}= n$\}}&\nonumber\\
    \label{eq:110}
    x^4 - \frac{3}{2}m x^3 - 2x^2 +3 m x + n &= 0.
  \end{align}
\end{subequations}
Solutions of this quartic equation lead to $\epsilon_0$. That is, for a given set of $w_0,w'$, it is
possible to get multiple solutions for $\epsilon_0$. Once $\epsilon_0$ is known, comparing the
coefficients of $R^1$ in~\eqref{eq:91} one can obtain $\epsilon_1$ as
\begin{subequations}
  \label{eq:111}
  \begin{align}
    \label{eq:112}
    f_1(\epsilon_0, \epsilon_1) &= \alpha_0 \beta_1 \epsilon_0-\frac{1}{C}  = 0\\
    \label{eq:113}
    \epsilon_1 &= \frac{1}{B \alpha_0 C \left(\frac{8}{9} B C_2 k_0^{-\frac{2}{3}} + \frac{2}{9} B
                 k_0^{-\frac{4}{3}} + \frac{4}{3} k_0^{-\frac{1}{3}} - \frac{4}{3} B C_2 \right)}.
  \end{align}
\end{subequations}
Finally, with $\epsilon_0, \epsilon_1$ known, $\epsilon_2$ can be obtained by comparing the
coefficients of $R^2$,
\begin{subequations}
  \label{eq:114}
  \begin{align}
    \label{eq:115}
    f_2(\epsilon_0,\epsilon_1,\epsilon_2) &=  \alpha_1 \beta_1 \epsilon_0 + \alpha_0 \beta_1 \epsilon_1 + \alpha_0 \beta_2 \epsilon_0 = 0\\
    \label{eq:116}
    \epsilon_2 &= \frac{\epsilon_1^2}{2 \epsilon_0} + \frac{\epsilon_0}{B}\frac{\frac{8}{27}B C_2 k_1^2
                 k_0^{-\frac{5}{3}} + \frac{4}{27}B k_1^2 k_0^{-\frac{7}{3}} + \frac{2}{9}k_1^2k_0^{-\frac{4}{3}} +
                 \beta_1 \left(\frac{\alpha_1}{\alpha_0} - \frac{\epsilon_1}{\epsilon_0}\right)}{- \frac{8}{9}B C_2
                 k_0^{-\frac{2}{3}} -\frac{2}{9}B k_0^{-\frac{4}{3}} +\frac{4}{3}B C_2 - \frac{4}{3}
                 k_0^{-\frac{1}{3}}}.
  \end{align}
\end{subequations}

\section{Condition for Jordan frame collapse in the presence of dust}
\label{sec:appendix2}
Here we argue that Jordan frame collapse is not possible in a spatially flat
universe consisting of dust and a quintessence field with logarithmic $w_\varphi$. We start with the
general condition for collapse in the presence of dust, with $\Omega_{K0} = 0$, from~\eqref{eq:84}
\begin{align}
  \label{eq:117}
  w_\varphi(\at) > \mathcal{C}(\at)
\end{align}
where,
\begin{subequations}
  \label{eq:118}
  \begin{align}
    \label{eq:119}
    w_\varphi(\at) &= w_0 - w' \ln \at, \\
    \label{eq:120}
    \mathcal{C}(\at) &= 1 + 2 \frac{\rho_m}{\rho_\varphi} \\
    \label{eq:121}
                   &= 1 + 2 \frac{\Omega_{m0}}{\Omega_{\varphi 0}}  \exp \left( 3 w_0 \ln \at - \frac{3}{2} w' (\ln \at)^2\right)\\
    \label{eq:122}
                   &= 1 + 2 \Omega_r \exp \left( 3 w_0 \ln \at - \frac{3}{2} w' (\ln \at)^2\right).
  \end{align}
\end{subequations}
Let us now consider the following arguments
\begin{enumerate}
\item It is obvious from~\eqref{eq:120} that $\mathcal{C}(\at)$ has a lower bound of $1$,
  $\mathcal{C}>1$. Since $w_\varphi(\at)$ is a monotonically increasing function of $\at$, $w_\varphi$ can intersect with
  $\mathcal{C}$ only after $w_\varphi$ crosses $1$. This happens at
  \begin{align}
    \label{eq:123}
    \at_* = \exp \left(\frac{w_0}{w'} - \frac{1}{w'} \right).
  \end{align}
  Thus, it is sufficient to consider only the range $\at>\at_*$ in order to check the possibility
  of the intersection.
\item The slopes of these functions are given by
  \begin{subequations}
    \label{eq:124}
    \begin{align}
      \label{eq:125}
      w_{\varphi,\at}(\at) &= \frac{- w'}{\at} \\
      \label{eq:126}
      \mathcal{C}_{,\at}(\at) &= 6 \Omega_r (w_0 - w' \ln \at) \exp \left( (3 w_0 - 1) \ln \at - \frac{3}{2} w' (\ln \at)^2 \right).
    \end{align}
  \end{subequations}
  Where the subscript $,\at$ denotes the derivatives of the functions with respect to $\at$.
\item At the point $\at_*$, the slope of $\mathcal{C}$ is higher than the slope of $w_\varphi$. To see this
  consider the ratio of the slopes
  \begin{align}
    \label{eq:127}
    \frac{\mathcal{C}_{,\at}}{w_{\varphi,\at}} = - \frac{6 \Omega_r}{w'} \left( w_0 - w' \ln \at \right) \exp \left( 3 w_0 \ln \at - \frac{3}{2} w' (\ln \at)^2 \right).
  \end{align}
  At the point $\at_*$, this ratio becomes
  \begin{align}
    \label{eq:128}
    \left.  \frac{\mathcal{C}_{,\at}}{w_{\varphi,\at}} \right|_{\at=\at_*} = - \frac{6 \Omega_r}{w'} \exp \left( \frac{3}{2 w'} (w_0^2 - 1) \right)
  \end{align}
  The observationally consistent range of $w_0$~\eqref{eq:15} is $-1.09<w_0<-0.66$. However we do not
  consider the possibility of $-1>w_0$ because this does not lead to a real $\varphi$~\eqref{eq:17}.
  Now for $-1<w_0<0$ and $w'<0$, the argument of the exponential is non-negative. The minimum possible
  value of the RHS is obtained when $w'$ is most negative and $\Omega_{mo}$ is smallest. Putting
  $w_0=-1,\ w'=-1.22, \Omega_r=.26/(1-.26)$ we get
  \begin{align}
    \label{eq:129}
    \left.  \frac{\mathcal{C}_{,\at}}{w_{\varphi,\at}} \right|_{\at=\at_*} = 1.73.
  \end{align}
  Thus we conclude that at $\at_*$, $\mathcal{C}$ has a higher slope than $w_\varphi$ given our choice of
  parameters.
\item In the range $\at>\at_*$, the slope of $w_\varphi(\at)$ always decreases. This is obvious since
  \begin{align}
    \label{eq:130}
    w_{,\at \at} = \frac{w'}{\at^2} <0.
  \end{align}
  On the other hand
  \begin{align}
    \label{eq:131}
    \mathcal{C}_{,\at \at} &= 6 \Omega_r \exp \left( (3 w_0 -2 ) \ln \at - \frac{3}{2} w' (\ln \at)^2 \right) \left[ 3 w'^2 (\ln \at)^2 + (w' - 6 w_0 w') \ln \at + 3 w_0^2 -w_0 - w' \right] \\
    \label{eq:132}
    \mathcal{C}_{,\at \at} &= 6 \Omega_r \exp \left( (3 w_0 -2 ) \ln \at - \frac{3}{2} w' (\ln \at)^2 \right) h(\at).
  \end{align}
  From this we see that the sign of $\mathcal{C}_{,\at \at}$ is always determined by the sign of
  $h(\at)$. We now argue that $h(\at)$ is always positive in the range $\at>\at_*$.

  To see this, first note that $h(\at)$ has a minimum at $\ln \at_1 = - \frac{1}{6w'} + \frac{w_0}{w'}$,
  \begin{align}
    \label{eq:133}
    h_{,\at}(\at) = \frac{w'}{\at} (1 - 6 w_0 + 6 w' \ln \at),
  \end{align}
  also,
  \begin{align}
    \label{eq:134}
    h_{,\at \at}(\at) &= \frac{w'}{\at^2} (-1 + 6 w_0 + 6 w' - 6 w' \ln \at)\\
    h_{,\at \at}(\at_1) &= 6 w'^2 \exp \left( \frac{1 - 6w_0}{3w'} \right) >0.
  \end{align}
  Also the point at which $h(\at)$ is minimum comes before the point $\at_*$, since
  \begin{subequations}
    \label{eq:135} 
    \begin{align}
      \label{eq:136}
      \frac{1}{-w'} + \frac{w_0}{w'} &> \frac{1}{-6w'} + \frac{w_0}{w'}\\
      \ln \left( \frac{1}{-w'} + \frac{w_0}{w'} \right) &> \ln \left(\frac{1}{-6w'} + \frac{w_0}{w'} \right)\\
      \at_* &> \at_1.
    \end{align}
    Thus in the range $\at>\at_*$, the minimum value of $h(\at)$ is the value at $\at_*$, which is
    \begin{align}
      \label{eq:137}
      h(\at_*) = 2 - w' > 0.
    \end{align}
  \end{subequations}
  Thus $h(\at) > 0$ in the range $\at > \at_*$, this implies $\mathcal{C}_{,\at \at} > 0$ in the
  range $\at > \at_*$. In words, the slope of the function $\mathcal{C}(\at)$ always increases in
  the range $\at > \at_*$.
\item Finally we note that for $0 < \at \leq \at_*$, $\mathcal{C}(\at) > w_\varphi(\at)$. At
  $\at_*$, $\mathcal{C}>w$ and $\mathcal{C}_{,\at} > w_{\varphi,\at}$. After $\at>\at_*$ the slope
  of $\mathcal{C}$ strictly increases while the slope of $w_\varphi$ strictly decreases. These
  functions hence can never intersect after $\at>\at_*$. Thus the condition $w_\varphi> \mathcal{C}$
  is never satisfied.

\end{enumerate}
\bibliographystyle{unsrt}
\bibliography{all_ref}
\end{document}